\documentclass[trackchanges]{aastex62}

\usepackage{blindtext}
\usepackage{graphicx}
\usepackage{soul}
\usepackage[utf8]{inputenc}
\usepackage{wrapfig}
\usepackage{threeparttable}
\usepackage{appendix}
\usepackage{hyperref}
\usepackage{placeins}
\graphicspath{{./}{}}
\received{XX}
\revised{XX}
\accepted{\today}
\submitjournal{ApJ}

\shorttitle{HMI Scattered Light}
\shortauthors{Norton et al.}
\begin{document}
\title{Stray Light Correction for the Helioseismic and Magnetic Imager}

\correspondingauthor{A.A. Norton}
\email{aanorton@stanford.edu}
\author[0000-0003-2622-7310]{A.A. Norton}
\affil{W.W. Hansen Experimental Physics Laboratory,
Solar Physics, Stanford University, Stanford, CA 94301-4085, USA}
\author{T.L. Duvall, Jr.}
\affil{W.W. Hansen Experimental Physics Laboratory,
Solar Physics, Stanford University, Stanford, CA 94301-4085, USA}
\affil{Max-Planck-Institut f{\"u}r Sonnensystemforschung, 
37077 Göttingen, Germany}
\author{J. Schou}
\affil{Max-Planck-Institut f{\"u}r Sonnensystemforschung, 
37077 Göttingen, Germany}
\author{R.S. Bogart}
\affil{W.W. Hansen Experimental Physics Laboratory, 
Solar Physics, Stanford University, Stanford, CA 94301-4085, USA}
\author{J. Zhao}
\affil{W.W. Hansen Experimental Physics Laboratory, 
Solar Physics, Stanford University, Stanford, CA 94301-4085, USA}
\author{C. Rabello-Soares}
\affil{W.W. Hansen Experimental Physics Laboratory, 
Solar Physics, Stanford University, Stanford, CA 94301-4085, USA}
\author{P.J. Levens}
\affil{W.W. Hansen Experimental Physics Laboratory, 
Solar Physics, Stanford University, Stanford, CA 94301-4085, USA}
\author{J.T. Hoeksema}
\affil{W.W. Hansen Experimental Physics Laboratory, 
Solar Physics, Stanford University, Stanford, CA 94301-4085, USA}
\author{C.S. Baldner}
\affil{W.W. Hansen Experimental Physics Laboratory, 
Solar Physics, Stanford University, Stanford, CA 94301-4085, USA}

\begin{abstract}
We report a point spread function (PSF) and deconvolution procedure to remove stray light from the Helioseismic and Magnetic Imager (HMI) data. Pre-launch calibration observations, post-launch Venus transit and lunar transit data were used to develop the PSF and evaluate how well it reproduced the observed scattering. The PSF reported differs from previous stray light removal efforts since we do not use Gaussians as the central mathematical component. Instead, we use a Lorenztian convolved with an Airy function. In 2018, the HMI team began providing full-disk, stray-light-corrected data daily. Intensity, Doppler, magnetogram, and vector magnetic field data are provided. The deconvolution uses a Richardson-Lucy algorithm and takes less than one second per full-disk image. The results, on average, show decreases in umbral continuum intensity, a doubling of the granulation intensity contrast, increases in the total field strength, most notably in plage by $\sim$1.4--2.5 the original value, and a partial correction for the convective blueshift. Local helioseismology analyses using corrected data yield results that are consistent with those from uncorrected data, with only negligible differences, except in sunspot regions. The new data are found in JSOC with names similar to the original but with the qualifying term '$\_dcon$'  or '$\_dconS$' appended, denoting whether the deconvolution was applied to the filtergrams or Stokes images. The HMI team recommends using the corrected data for improved visual clarity, more accurate irradiance reconstruction, better co-alignment with high-resolution data, reduced errors in tracking algorithms, and improved magnetic field strengths. 
\end{abstract}

\section{Introduction}\label{sec:intro}
Stray light, also known as parasitic light or scattered light, is a term used to describe light observed within an image element that does not originate from the corresponding location in the target being imaged, in this case, the solar disk.  In other words, it is light deflected from a straight path. Levels of stray light are reduced when observing with space-based instruments because the Earth's atmosphere is no longer an influence, but stray light is not altogether removed.  Optical systems can have aberrations of varying degrees and small imperfections (e.g., dust on an optical element) that cause scattering.  Stray light may vary spatially over the field of view and temporally as the optics age. The scattering can be  described by the point spread function (PSF), its Fourier transform, known as the optical transfer function (OTF) and the magnitude of that, known as the modulation transfer function (MTF).

%When characterizing the PSF of the instrument, there is the unavoidable blurring of the point source by the optics, i.e., the feature in the image that is regarded as the unresolved target, and the scattering from imperfections in the optics or from physical processes that occur in the solar atmosphere between emission and the telescope front window.  
The observed image, $O$, is modeled by a PSF, $\Psi$, convolved with the solar image $I$ as described by Equation \ref{Eq-1}.  
   \begin{eqnarray}   
     O(r,\phi) &=& \Psi(r,\phi) \ast I(r,\phi)           
      \label{Eq-1}    
   \end{eqnarray}

The PSF of an ideal optical system without scattering is best modeled by an Airy function \citep{airy:1838} as it describes the diffraction and interference pattern produced by a point source of light observed by a circular aperture of finite diameter. The PSF of a realistic system can be described as an Airy function convolved with a Lorentzian as the Lorentzian provides a description of the scattering. However, historically, a more common approach has been to model $\Psi$  using the sum of a Gaussian for the blurring component (the higher amplitude central peak of the function in frequency domain) and a dispersion function or a Lorentzian for the scattering component (the lower amplitude tail that accounts for low-level scattering from further away) \citep{zwaan:1965, jefferies:1991}, see Equation \ref{Eq-2}.  
   \begin{eqnarray}   
      O(r,\phi)  &=& I(r, \phi) \ast (\Psi_{b}(r, \phi) + \Psi_{s}(r, \phi))
     \label{Eq-2}
    \end{eqnarray}
For example, \citet{pierce:1977} propose a PSF in the form of Equation \ref{Eq-3}
   \begin{eqnarray}   
     \Psi(r) &=& ( 1 - \epsilon )e^{-(r/w)^2} +  \frac{\epsilon} {1 + (r/W)^\kappa}   
       \label{Eq-3}    
   \end{eqnarray}
while \citet{jefferies:1991} used a sum of two Gaussians for the blurring function and a dispersion function, applying this technique to ground-based K-line helioseismology data taken from the South Pole. 
The technique of using sums of Gaussians to remove stray light has been advanced considerably in the past twenty years by \citet{mathew:2007} who used the sum of three Gaussians and a Lorentzian applied to data from the Michelson Doppler Imager \citep{scherrer:1995}, by \citet{mathew:2009} who used four Gaussians applied to data from Hinode Solar Optical Telescope (SOT) \citep{tsuneta:2008}, and by \citet{yeo:2014} who applied a sum of five Gaussians to the data from HMI \citep{scherrer:2012}.

Another approach is reported by \citet{wedemeyer:2008} who derives a set of PSF for the Hinode SOT by the convolution of ideal diffraction-limited PSFs and Voigt functions. Gaussians, although versatile and easier to handle mathematically than the Airy function, are not ideal to use in the PSF.  \citet{wedemeyer:2008} points out that the approach shown in Equations~\ref{Eq-2}-\ref{Eq-3} using a linear combination of functions is, strictly speaking, not correct. Instead the convolution method is preferable. Rather than one unique solution, there are multiple combinations of parameters, or a range of parameters, that fit the observations.  \citet{diazbaso:2018} developed a method to enhance HMI data by deconvolving and super-resolving, by a factor of two, single images using a neural network trained on synthetic data. Their resulting images are limited field-of-view and mimic observations with a diffraction-limited telescope twice the diameter of HMI. This work is of merit but the HMI team decided to use a more conservative approach to describe the instrumental PSF and create full-disk, deconvolved images. Note that the PSFs developed by \citet{wedemeyer:2008, yeo:2014, diazbaso:2018} do not account for scattered light from a distance greater than ten arcseconds away, which is approximately 20 HMI pixels, and only a third of the radius of Venus as observed during its 2012 transit.  

Therefore, we develop a PSF based on the convolution of a Lorentzian with an Airy function, as described in part in \citet{couvidat:2016}. The development of the PSF is discussed in Section~\ref{sec:sec2}, including the mathematical and optical specifics in Section~\ref{sec:sec2.1} and the fitting of the parameters based on observations in Sections~\ref{sec:sec2.2} - \ref{sec:sec2.4}.  The deconvolution procedure as applied to HMI data and the resulting data products are described in Section~\ref{sec:sec3}, with more details contained in the Appendix.  Changes in the science data are found in Section~\ref{ref:sec4} with a discussion of the scientific implications in Section~\ref{ref:sec5}. 

\section{Characterizing the Point Spread Function} \label{sec:sec2}
\subsection{Mathematical and Optical Specifics}\label{sec:sec2.1}
For the optics of an incoherent imaging system such as HMI, the intensity of the ideal PSF can be described as an Airy function which is proportional to the Bessel function of the first kind, J1, to give a non-negative intensity as shown in Equation 4 and \ref{Eq-4and5}.   
\begin{eqnarray}   
     PSF(r\prime) &=& \left( \frac {2J_1(r^{\prime})}  {r^\prime} \right)^2\\
     r^\prime &=& \frac{\pi Dr}{f \lambda}  
       \label{Eq-4and5}    
   \end{eqnarray}
where $r^{\prime}$ is a normalized radius. $D$ is the diameter of the telescope aperture, which is 14 cm for HMI.  $f$ is the effective focal length which is 4953 mm and $\lambda$ is 6173 \AA.  Each HMI pixel is 12 microns, so the $r^{\prime}$ value has this as an incremental value.   The first minimum of the Airy function is found at 1.22 $\lambda/D$ where $\lambda$ is the wavelength of the instrument and $D$ is the diameter of the aperture.  

\begin{figure}[htbp]
%\centerline{\includegraphics[width=0.85\textwidth,clip=]{Fig1_withyeo.png}}
\centerline{\includegraphics[width=0.75\textwidth,clip=]{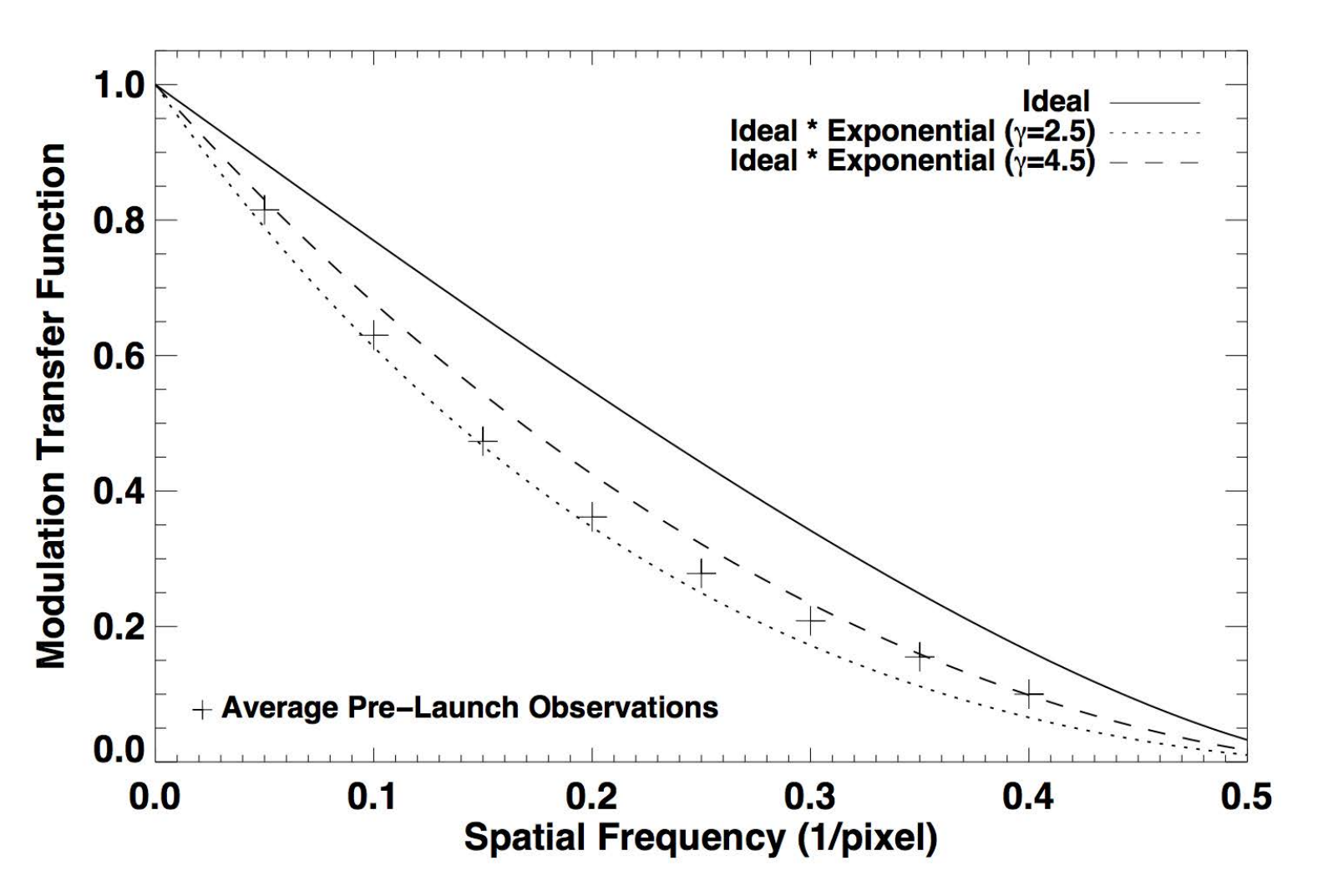}}
\caption{Several MTFs are shown as a function of spatial frequency.  The $+$ symbols represent the average of three pre-launch (i.e. ground-based) observations during instrument calibration,  as reported in \citet{wachter:2012} text and in their Figure 3. The ideal MTF, described by Equation~\ref{Eq-8}, is plotted as a solid line while the ideal MTF multiplied by exponentials with $\gamma$ values of 2.5 and 4.5 are also plotted as dotted and dash lines, since these two curves bracket the pre-launch observations that are shown as symbols. }
\label{fig:Fig1}
\end{figure}
It is worth mentioning that HMI is under-sampled, meaning that the size of the pixels are slightly larger than the size that would match the diffraction limited system. The reason for the under-sampled system  is that the instrument was initially designed for the 6768 \AA~Ni-I line.  When the spectral line was changed to be 6173 \AA~Fe-I to take advantage of the larger $g$-effective of the Fe-I line, the aperture size was not adjusted from $\sim$14 to $\sim$12 cm.  The optimal size of HMI pixels to be diffraction limited would be 0.45$^{\prime\prime}$, not $\sim$0.50$^{\prime\prime}$ which is the actual HMI pixel size.  The effect of the undersampling is that there is some power in the MTF above the spatial Nyquist frequency, as can be seen in Figure~\ref{fig:Fig1} since the MTF does not go to zero at the 0.5 spatial frequency. 

The Optical Transfer Function (OTF) is described by Equation 6 as found in \citet{bracewell:1995}.   The OTF is the Fourier transform of the PSF.  Equations~\ref{Eq-6}-\ref{Eq-8} represent the ideal OTF which is the Fourier transform of the Airy function, a normalized spatial frequency and the ideal MTF.  In older texts, Equation~\ref{Eq-6} was often referred to as the chat function, but is now usually referred to as the normalized OTF.  
\begin{eqnarray}   
     OTF(\rho^{\prime}) &=& \frac{2}{\pi} [acos(\rho^{\prime}) -\rho^{\prime}\sqrt{(1-\rho^{\prime2})}] \label{Eq-6}\\
     \rho^\prime &=& \frac{\rho f \lambda} {D}  \label{Eq-7}\\
     MTF_{ideal}(\rho\prime) &=& |OTF(\rho\prime)| 
       \label{Eq-8}    
   \end{eqnarray}
In the above Equations \ref{Eq-6}-\ref{Eq-8}, $\rho$ is the spatial frequency, $\rho\prime$ is a normalized spatial frequency, and all other symbols are explained above. The MTF is the absolute value of the OTF as described by Equation~\ref{Eq-8} and plotted as the solid line in Figure 1.  

The realistic form of the MTF is shown in Equation \ref{Eq-9} which is the real part of the OTF multiplied by an exponential. The Fourier transform of an exponential is a Lorentzian, so the form in Equation \ref{Eq-9} represents the Fourier transform of a realistic PSF -- an Airy convolved with a Lorentzian in the image plane. The range of values explored for $\gamma$ were 2.5-4.5 since this range was determined by the ground-based observations. The curves for Equation~\ref{Eq-9} with $\gamma$ of 2.5 and 4.5 are seen in Figure~\ref{fig:Fig1}.  
\begin{eqnarray}   
    MTF(\rho^{\prime}) &=& |OTF(\rho^{\prime})| \times e^\frac{ -\pi\rho^\prime} {\gamma}  \label{Eq-9}
\end{eqnarray}    
The final value of $\gamma$ in the exponential function was constrained by the pre-launch calibration data (see Section 2.2), and determined through least-squared fitting of the Venus transit data (see Section 2.3).  

\begin{figure}[b]
   \centerline{\hspace*{0.015\textwidth}
    \includegraphics[trim=30 10 30 30, clip,width=0.65\textwidth,clip=]{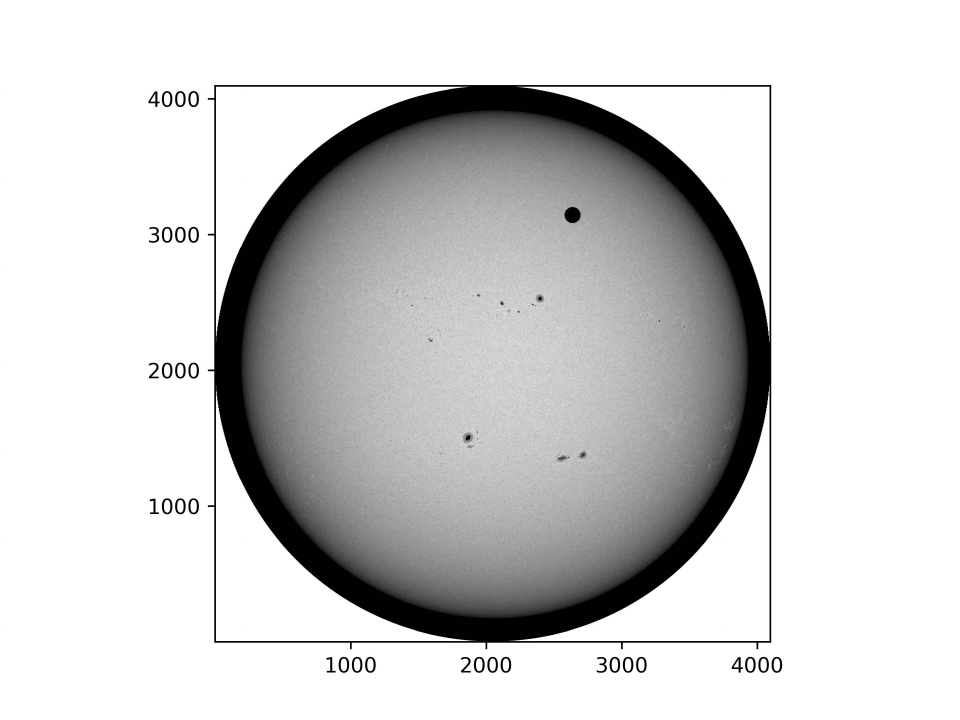} }
\caption{A single filtergram image from the HMI side camera taken during the transit of Venus on 2012.06.06 at 02:04 UT is shown with pixels labelled on the x and y axis. Several sunspots are visible as well as the disk of Venus. A blank image with the same radius and center with limb darkening was generated with a circle of zeroes in the position of Venus. The blank image represented a solar image without any scattered light; the light level off the limb and within the disk of Venus is zero.  A guess PSF was convolved with the blank image to estimate the scattered light distribution in a forward modeling process.  This process was iterated until the result of the forward model best matched the observed image.}
\label{fig:Fig2}
\end{figure}

\subsection{Ground-based Calibration Using Targets and Field Stops}\label{sec:sec2.2}

\citet{wachter:2012} reported on ground-based testing of HMI optics using a combinations of targets (random dot, star, star array, and field stop) to compare observed spatial power with those of the known targets. They report the modulation transfer function measured in two directions. The field stop target had a radius of 400 pixels, meaning that long-range scattering was measurable outside of that field stop. They report an HMI scattered light PSF using the form of Equation~\ref{Eq-3} with the parameters $W=3.0$, $\epsilon=0.1$, $\kappa=3.0$, $w=1.8$ as the best fit.  The long-range scattered light level is captured by this PSF. However, it is noted that the while ``the overall fit is good, a small deviation near the edge can be attributed to the simplified representation of the core function as a Gaussian." Ground-based calibration measurements were taken on at least three different dates during which thermal and vibrational conditions varied.  The scatter measured in the azimuthally averaged MTF from ground observations was on the order of +/- .0125 at the larger spatial frequencies. Although this might motivate testing a value of $\gamma$ slightly larger than 4.5, Venus transit data do not support higher values of $\gamma$. The observed ground-based values are plotted in Figure~\ref{fig:Fig1}. Our desire to improve the form of the PSF by using an Airy convolved with a Lorentzian motivated us to
simply take the observed MTF values reported in \citet{wachter:2012} to constrain the range of values of $\gamma$ in our exponential component. Figure~\ref{fig:Fig1} shows the the ideal MTF, 
%an example of the sum of Gaussians approach by \citet{yeo:2014}, 
and two MTFs that are the ideal convolved with the exponentials using $\gamma$ values of 2.5 and 4.5 that bracket the observed values.

 \begin{table*}[hb]
\caption{Data used for PSF development\label{tab:dataproducts}}
\centering
\begin{tabular}{llll}
\hline  
Data product & Time & Camera & Description\\
\hline
\hline
hmi.lev1\_cal & 2012.06.05\_22:09 $-$ 06\_04:49 &Side & Filtergrams, FID=10004, Transit of Venus  \\
%was 06.06_02:04-02:46
hmi.lev1 & 2010.10.07\_11:47:00& Front & Filtergram, FID=10159, Lunar Transit \\
hmi.lev1 & 2016.07.14\_13:48:00 & Both & All filtergrams in 720 seconds\\
hmi.S\_720s & 2016.07.14\_13:48:00 & Both & Average Stokes polarization images\\
\hline 
\end{tabular}
\label{tab:table1}
\end{table*}

\subsection{Transit of Venus Data}\label{sec:sec2.3}

The transit of Venus occurred on 2012 June 5-6.  Ephemeris data were calculated to determine the center of the Venus disk in relation to the HMI field of view and Sun frame position.  HMI possesses two cameras: the front camera (also known as camera 2 or Doppler camera) and the side camera (aka camera 1 or the magnetic camera).   During the Venus transit, the standard observable sequences were taken on the front camera, and a series of continuum images in linear polarization were taken on the side camera.  Therefore, we used data from the side camera. The continuum images were recorded on the side CCD in the continuum 0.344$\,${\AA} from line center at rest.  These observations were taken at a 3.75 second cadence during the Venus transit. An example of this data is shown in Figure~\ref{fig:Fig2}.  
%The ephemeris data file can be located within JSOC.\url{http://jsoc.stanford.edu/SUM15/D330144691/S00000/SOLAR_TRANSIT_2012157.S02}.
%\url{http://jsoc.stanford.edu/SUM15/D330144691/S00000/SOLAR_TRANSIT_2012157.S02}
The radius of Venus was 0.00803$^{\circ}$, or 29.5$''$, or 58.5 pixels with an HMI pixel scale of 0.504$''$ per pixel.  We assume that Venus is a perfect sphere. Table~\ref{tab:table1}, row 1 lists the Venus transit data used in the PSF determination. 

As part of the general data processing, the HMI images are corrected for spatial distortion \citep{wachter:2012}. The distortion is caused by focus, temperature and various optical inhomogeneities. Distortion maps are calculated separately for the side and front camera and applied to the individual filtergrams.  Although the distortion correction is quite small, it is important to remove it prior to characterizing the scattered light. As such, the PSF determined from our efforts is meant to be applied only to data in which spatial distortion has been removed.

\begin{figure}[b]
  \centerline{\hspace*{0.015\textwidth} 
            \includegraphics[width=0.80\textwidth,clip=]{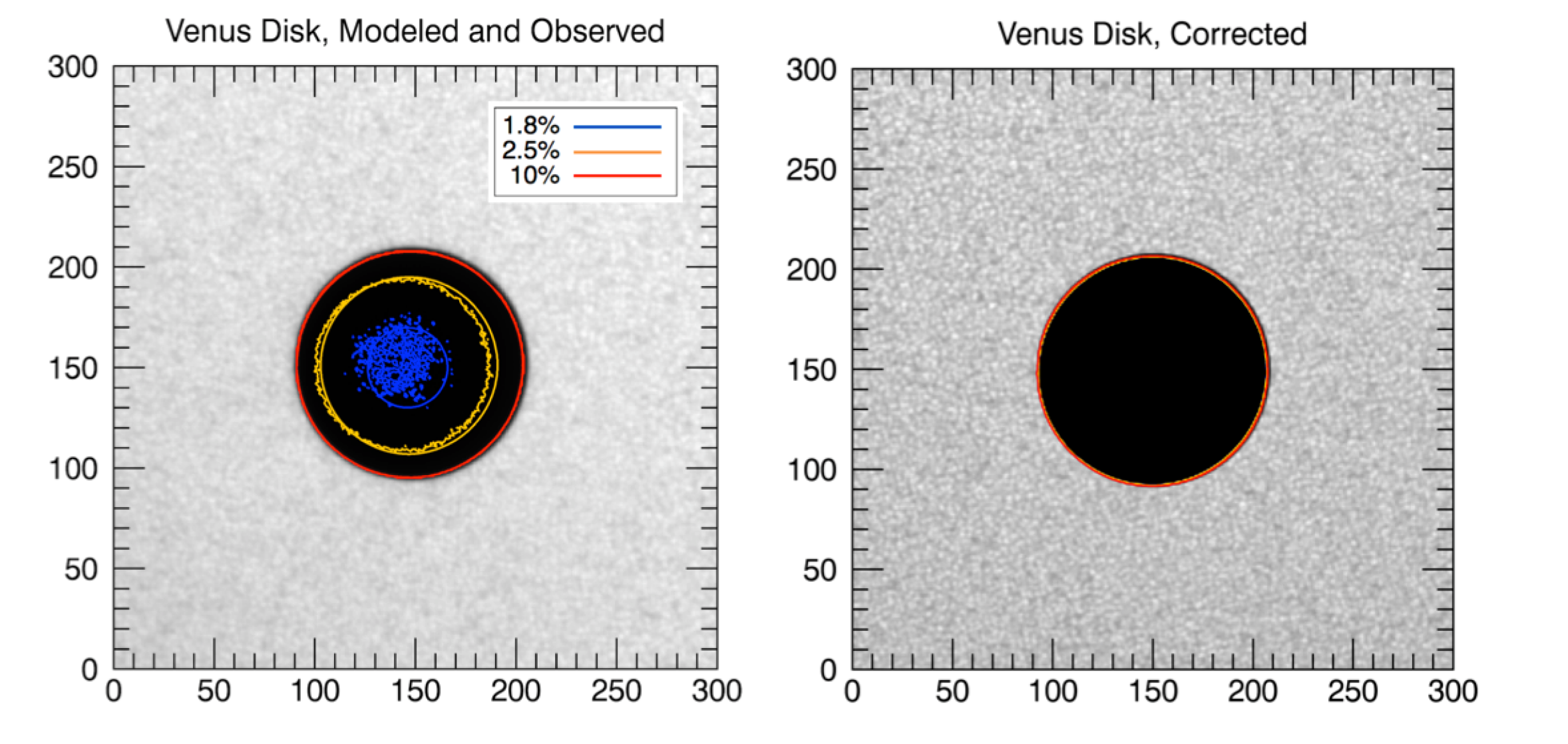}  }
\caption{The disk of Venus is shown with the observed and forward modeled light level contours overplotted. The observed contours appear as irregular circles, while the forward-modeled contours are smooth circles. The forward model uses a disk of zeros placed within a limb-darkened solar disk that is convolved with the PSF. The azimuthal asymmetry is apparent. We do not include an azimuthal dependence in the form of the PSF even though it is known to exist. 
}
\label{fig:Fig3}
\end{figure}

The total light level in the disk of Venus changes as the transit progresses due to granulation differences and $p$-mode oscillations contributing to the scatter in individual filtergrams. We found that the most stable light levels in the disk of Venus during the transit were between 02:04-02:46 on 6 June 2012, as was also reported by \citet{yeo:2014}. These data can be found in JSOC denoted as hmi.lev1\_cal[2012.06.06\_02:04/42m][?camera=1?][?FID=10004?]. We selected four filtergrams to span this forty-minute period in the transit. The filter identification numbers of FID = 10004 represent the $I+Q$ polarization combination.   
 
There may not be a unique solution to the PSF for HMI, meaning that several combinations of parameter values, or a range of values for a given parameter, may perform equally well at recreating the observed scattered light levels. We found this to be the case in that a range of values used in the PSF during deconvolution resulted in a disk of Venus that contained no scattered light.  Therefore, forward modeling needed to be used for determining the PSF. 

In order to forward model the scattered light, we created artificial images of the Sun with the same radius and center and with limb darkening. A circle with the radius of Venus containing intensity values of zero was inserted in the position of Venus. This artificial image represents a solar image without any scattered light; the light level off the limb and within the disk of Venus is zero.

\begin{figure}[hb]
  \centering
  \includegraphics[trim=55 30 245 515, clip, width=.4\linewidth]{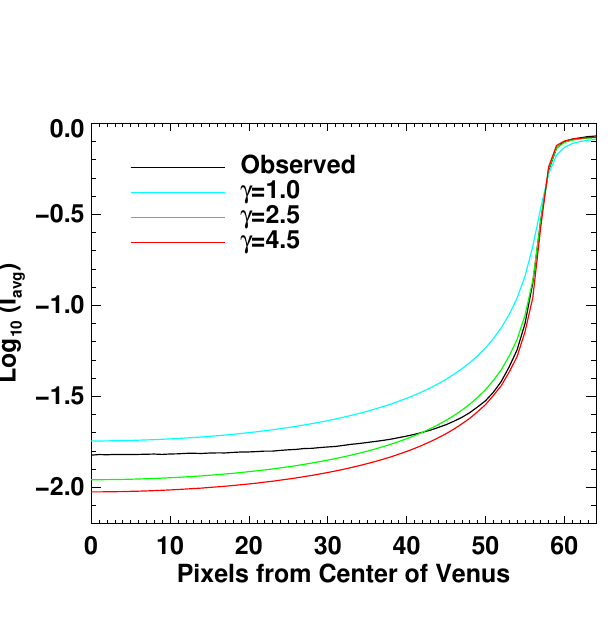} \qquad
  \includegraphics[trim=55 30 245 515, clip, width=.4\linewidth]{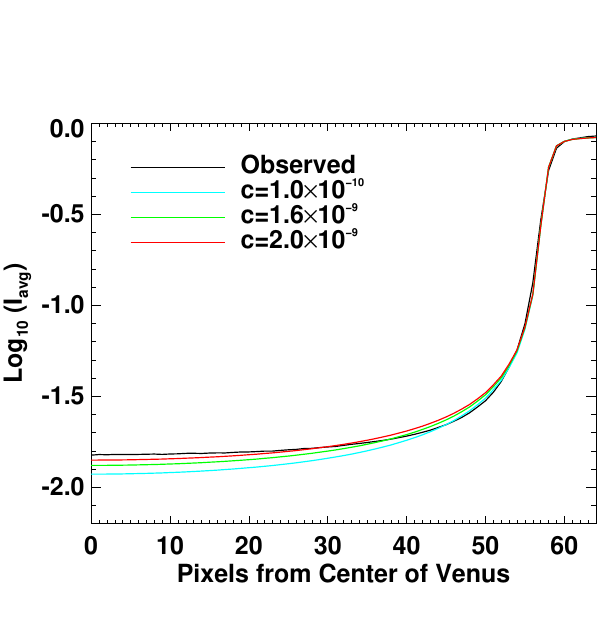} \qquad
  \includegraphics[trim=55 20 140 515, clip, width=.5\linewidth]{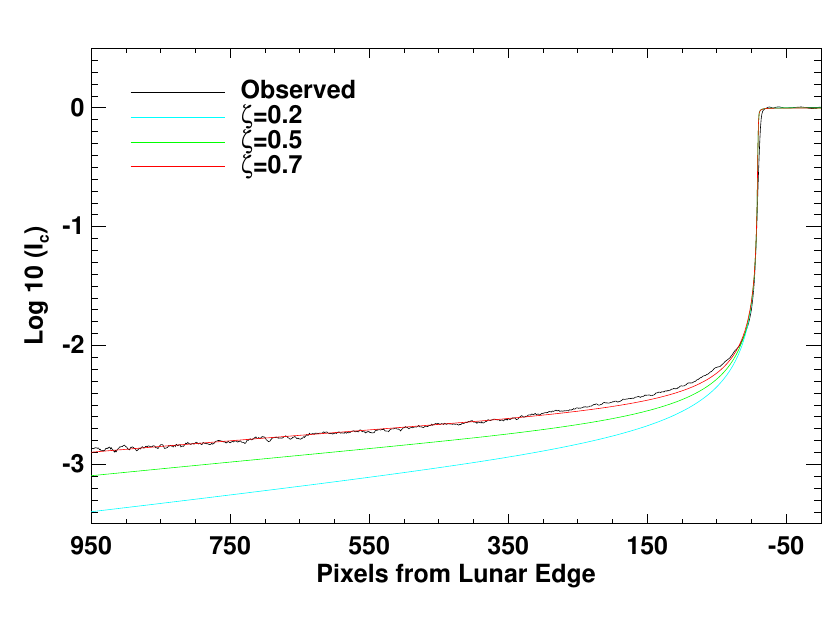} \qquad
  \caption{The scattered light levels dependence on $\gamma$, $c$, and $\zeta$ as shown in Equations 9-10. The azimuthally averaged, normalized intensity observed in the disk of Venus (black line) is plotted on logarithmic scale as a function of pixels from the center of Venus in the top two plots and from the lunar edge during a lunar eclipse in the bottom panel. Forward modeled intensities are calculated by convolving a PSF with artificial data and light levels for different values of $\gamma$ (top left), $c$ (top right) and $\zeta$ (bottom). For the top left panel, $c$ and $\zeta$ are 0. For the top right and lower panels, a $\gamma$ of 4.5 is used, and a $c$ of 2$\times$10$^{-9}$ is used for the lower panel. Less light is scattered into the center of Venus with increasing values of $\gamma$ since the FWHM of its corresponding Lorentzian is proportional to $\frac{1}{\gamma}$. The c value is a constant multiplied by exponential term and simply raises the light level far away from any source. }
  \label{fig:Fig4}
\end{figure}
Observed light levels in the center of the disk of Venus were 1.8\% of the nearly quiet-Sun continuum intensity, see the left panel in Figure~\ref{fig:Fig3}, with three light level contours shown.  The forward-modeled light level contours are plotted over the observed light level contours.  The parameters in the guess PSF were iterated until the difference between the observed and forward modeled values were minimized, although there was some redundancy in the minimization.

This PSF was used to deconvolve the full disk image. The resultant, corrected Venus disk is shown on the right panel of Figure~\ref{fig:Fig3}, where scattered light has effectively been removed so that all three light level contours lie at the very edge of the disk of Venus.  An azimuthal asymmetry clearly exists as seen in the observed contours being shifted to the left and upward in relation to forward-modeled contours in the left panel of Figure~\ref{fig:Fig3}. For simplicity, we chose not to include an azimuthal component in our form. 

 However, ignoring the azimuthal asymmetry may have implications. Since the scattering from bright surroundings into dark features can be slightly stronger along some position angles than others, deconvolution with a symmetric PSF may (a) over-correct along some directions and (b) under-correct along others. Because the asymmetry is most pronounced at faint intensity levels, it is consistent with an effect in the PSF wings (far wings/tail) rather than in the core. Consequently, the main impact is on narrow structures and edge sharpness: penumbral filaments may appear marginally sharper along one position angle than another, and umbral minima may differ by a small fraction of $I_c$ depending on the local brightness geometry. In plage, any residual anisotropy should be seen as very small changes to contrast at the periphery, which can subtly affect apparent or computed areas of the features. 

The criteria for successful forward modeling was to be within $\sim$100 counts of the true observed image, equivalent to 0.1\% of the continuum intensity on the solar disk.  The forward modeled levels are compared to observed levels also in  Figure~\ref{fig:Fig4} top panels, with light levels plotted on a log scale such that the difference between the observed line ($log_{10}(I_{c}) = - 1.82$) and the best-fit modeled line ($log_{10}(I_{c}) = -1.84$) near the center of Venus is equivalent to 0.07\% . 

\subsection{Lunar Transit Data}\label{sec:sec2.4}

Lunar transits, where the Moon obscures a portion of the solar image as observed from the SDO spacecraft, occur between one and four times a year. These offer an opportunity to observe the scattered light levels far away from the solar disk.  One such transit occurred on 2010 October 07. Table~\ref{tab:table1}, row 2 lists the lunar transit data used in the PSF determination.  The light levels are roughly 1\%, 0.3\% and 0.1\% at distances 10$^{\prime\prime}$, 100$^{\prime\prime}$ and 700$^{\prime\prime}$ away from the limb, see Figure~\ref{fig:Fig4},  lower panel.  Similar to description in \S2.3, an artificial image of the lunar transit was created using ephemeris data for Moon center and size as well as solar limb darkening.  This artificial image was then used to test any given PSF by convolution and comparing the resultant light curves with the observed lunar transit image. 

Analysis from the lunar transit and off-limb data showed that the light level continued to be small but positive with with increasing distance from the solar limb (see Figure~\ref{fig:Fig4}, lower panel). This motivated adding an additional term to the PSF, see term 2 in  Equation~\ref{Eq-10}, which when included, provides a better fit for light levels at a distance of 40 Mm (100 pixels) or greater, see Figure~\ref{fig:Fig4}, lower panel. This form of the PSF proved adequate for our purposes.

\begin{eqnarray}
    PSF(r^{\prime}) &=& \mathcal{F}(MTF) + c\times e^\frac{-\pi r^{\prime}}{\zeta}
    \label{Eq-10}    
\end{eqnarray}
  
The final values and terms of Equation~\ref{Eq-8} (MTF) and Equation~\ref{Eq-10} (PSF) are: $\gamma$ of 4.5, $\rho$ and $\rho^{\prime}$ are the spatial frequency and normalized spatial frequency, ${\mathcal{F}}$ denotes the Fourier transform, $c$ is a constant of $1.54 \times 10^{-9}$, $\zeta$ is equal to 0.7, and $r^{\prime}$ is the spatial distance from the center of the image in pixels, normalized by 2048. See Figure~\ref{fig:Fig4} for light level estimates in the Venus disk and during the lunar transit as determined from various values of $\gamma, c$, and $\zeta$.

\section{Deconvolution as Applied to HMI Data}\label{sec:sec3}

We use the Richardson-Lucy algorithm for deconvolution {\citep{richardson:1972,lucy:1974}}. This is an iterative method that returns a maximum-likelihood solution for the original image.  It is popular because it preserves the total number of counts of the original image and returns non-negative values. Whereas convolution is straightforward because it is multiplication, deconvolution can be problematic, basically because it is division, so any small values in the frequency domain, due to low signal-to-noise ratio in that frequency range, can result in amplification in the resulting signal. Increased noise in the low-signal data regions is common, and Gibb's phenomenon, a type of ringing at the edges of features, occurs. Limiting the number of iterations, considered a regularization strategy, can reduce the noise introduced at high spatial frequencies.  We limit the number of iterations to twenty. 

The deconvolution can be applied to either the individual filtergrams or the Stokes images.  
%\subsection{Individual Filtergrams to Produce 45-second Data}
To create the 45-second data, we deconvolve the PSF for each filtergram taken every 3.75 seconds.  The code is written in C++, runs in the JSOC data processing environment, is implemented after the flat field and dark are applied, and after distortion is removed, but before the filtergrams are combined to produce the science data.  The new data products associated with the 45-second data are shown in the first five rows in Table~\ref{tab:A1} in the Appendix, denoted by appending '\_dcon' to the original data product names. 
%\subsection{Averaged Stokes Images to Produce 720-s Data}

Alternatively, one can significantly reduce the amount of computing time needed by applying the deconvolution to the six Stokes polarization images, [I$\pm$Q, I$\pm$U, I$\pm$V], for the six wavelength positions across the spectral line. This approach is applied to the 720-second data. The new data products associated with the 720-second data are shown in Table~\ref{tab:A1} in the Appendix, denoted by appending '\_dconS' to the original product names. All of the '\_dcon' and '\_dconS' data products are publicly available through JSOC.    

\section{Changes to Science Data}\label{ref:sec4}
We now compare the original and scattered light corrected data from the commonly produced HMI data products to quantify changes in the intensities, magnetic field quantities, Doppler velocities, and helioseismic inversions. 

\subsection{Intensities}

One way to summarize the changes in the I$_c$ data is simply that the scattered light correction causes dark features to become darker and bright features to become brighter.  To further examine the intensity changes, the `true continuum' data is analyzed. 

\begin{figure}[b]
  \centering
  \includegraphics[trim=40 10 60 0, clip, width=.34\linewidth]{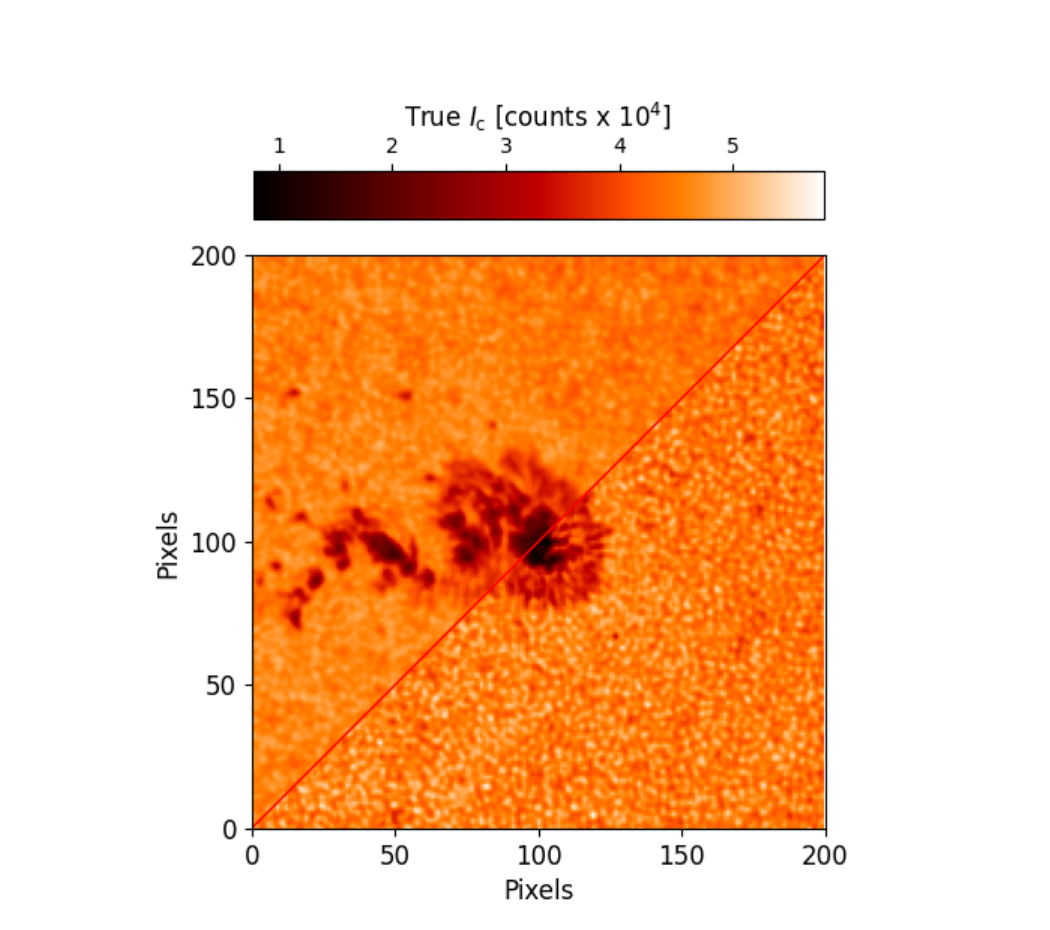} 
    \includegraphics[trim=40 0 60 40, clip, width=.28\linewidth]{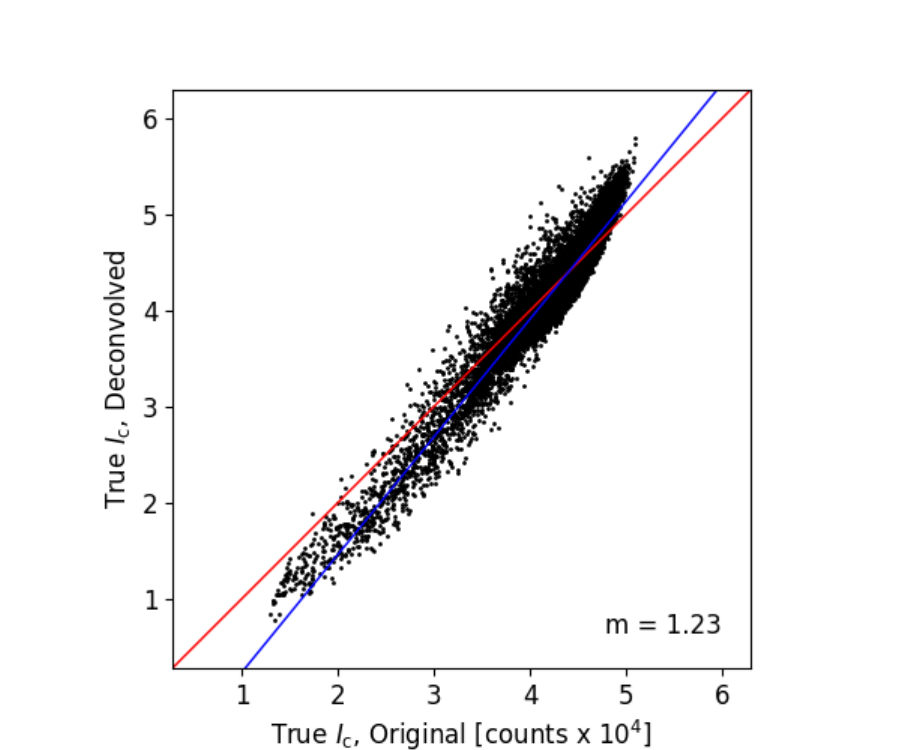} \qquad
  \includegraphics[trim=40 0 60 40, clip, width=.28\linewidth]{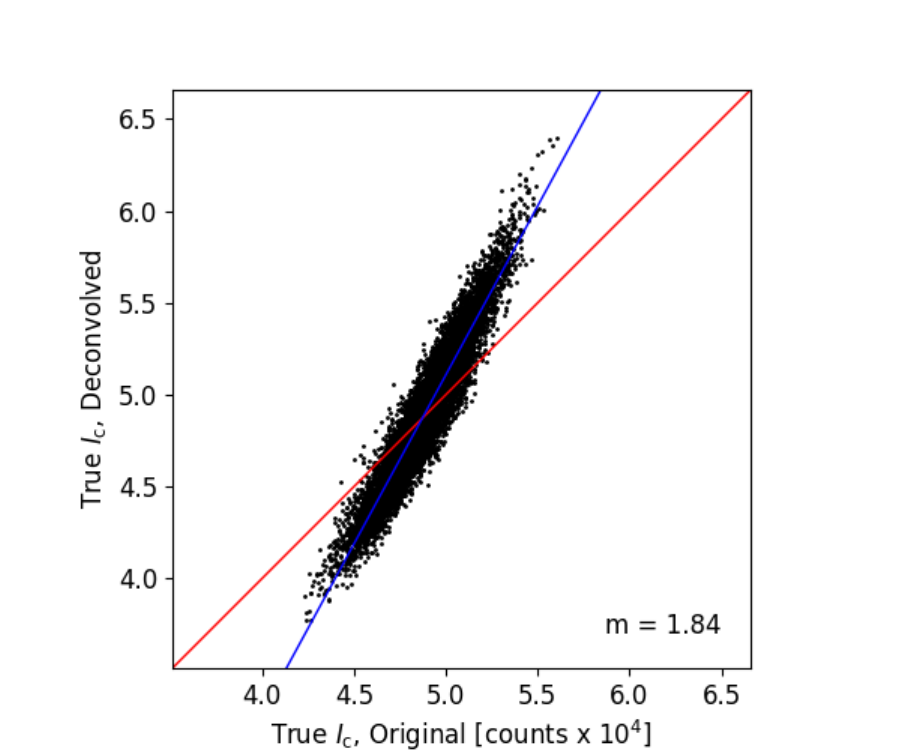} \qquad
  \caption{Comparisons are shown for the true continuum from 2022.12.31 06:00 UT (NOAA 13179, HARP 8927) with the plot on the left showing image data for the original (upper, left) and deconvolved data (lower, right). Scatter plots of the true continuum values are shown with the original versus deconvolved data for the 200 $\times$ 200 pixel region for active region and penumbra pixels (middle) and quiet-Sun pixels (right). The slope value, m, is shown in the panel as corresponding to a linear fit (blue) with a unity line (red) for context. }
  \label{fig:Fig5}
\end{figure}

The HMI measure of `true continuum' are filtergrams taken four times a day, two each at the time specified as T\_REC of 06:00 UT and 18:00 UT, when the instrument is tuned 0.345 \AA~from line center. The true continuum data can be found by querying JSOC for hmi.lev1\_cal[?FID=10001?], the original data, or hmi.cont\_dcon, the stray-light corrected data. A comparison of the values for a 200 $\times$ 200 pixel field of view from NOAA 13179 observed on 2022.12.31 at 06:00 UT is plotted in Figure~\ref{fig:Fig5}. The stray-light corrected quiet-Sun data are darker in the intergranular lanes and brighter in the convective cell centers, see scatter plot in the right panel of Figure~\ref{fig:Fig5} with the slope of $m=1.84$ for the quiet-Sun pixels. The correction is less drastic in the magnetic field regions of umbra and penumbra, see scatter plot in the middle panel of Figure~\ref{fig:Fig5}, with a slope of $m=1.23$. Thresholds in intensity and magnetic field  were used to determine whether pixels were considered quiet-Sun or active region pixels. 

The reason the correction is more dramatic in quiet-Sun is that granulation causes nearby pixels to have more variation in their brightness values than in sunspots, i.e., the spatial gradient of the intensity contrast is steeper.  

\begin{figure}[t]
  \centerline{\hspace*{0.015\textwidth}
              \includegraphics[width=0.35\textwidth,clip=]{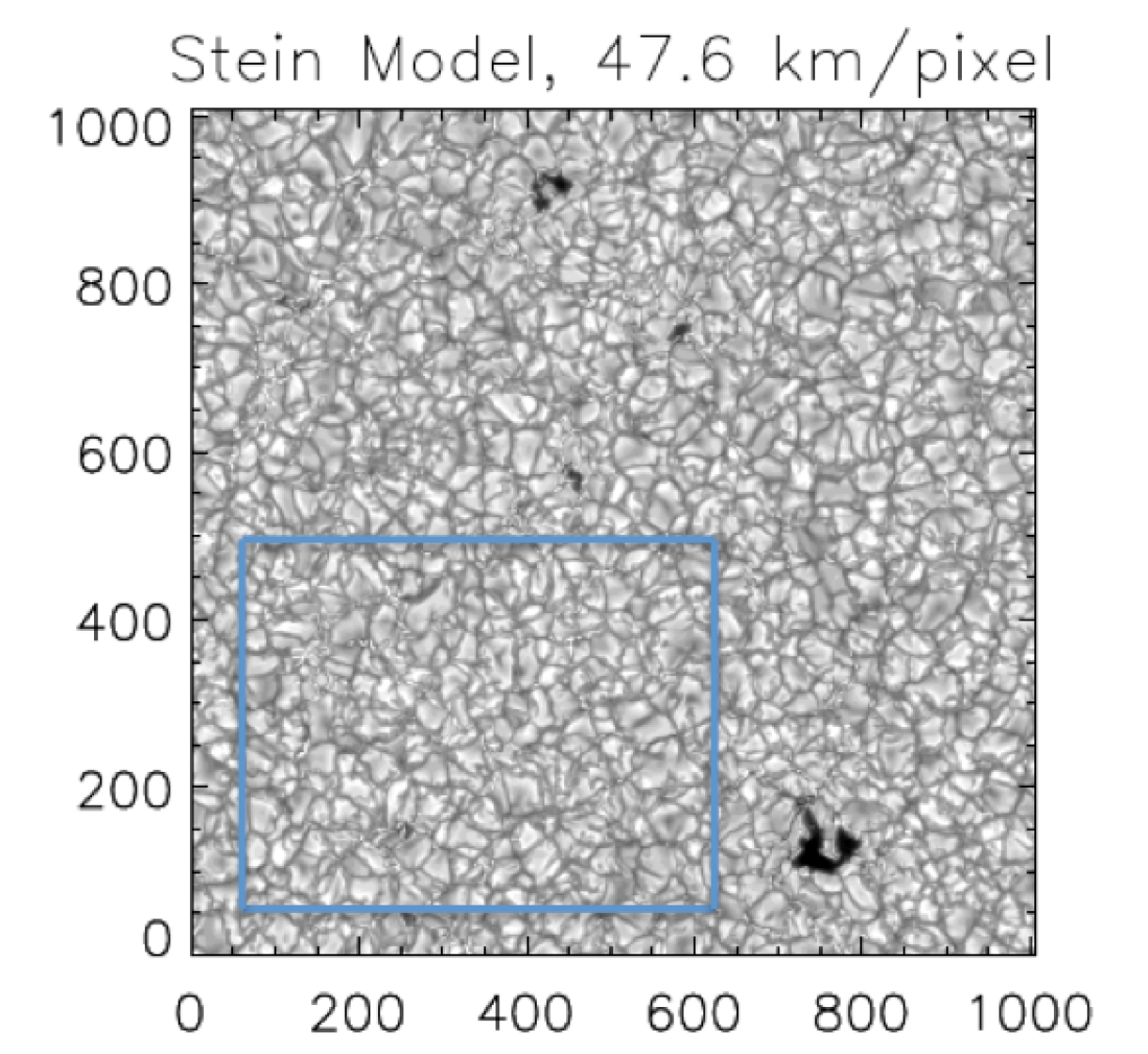}
            \hspace*{-0.03\textwidth}
               \includegraphics[width=0.35\textwidth,clip=]{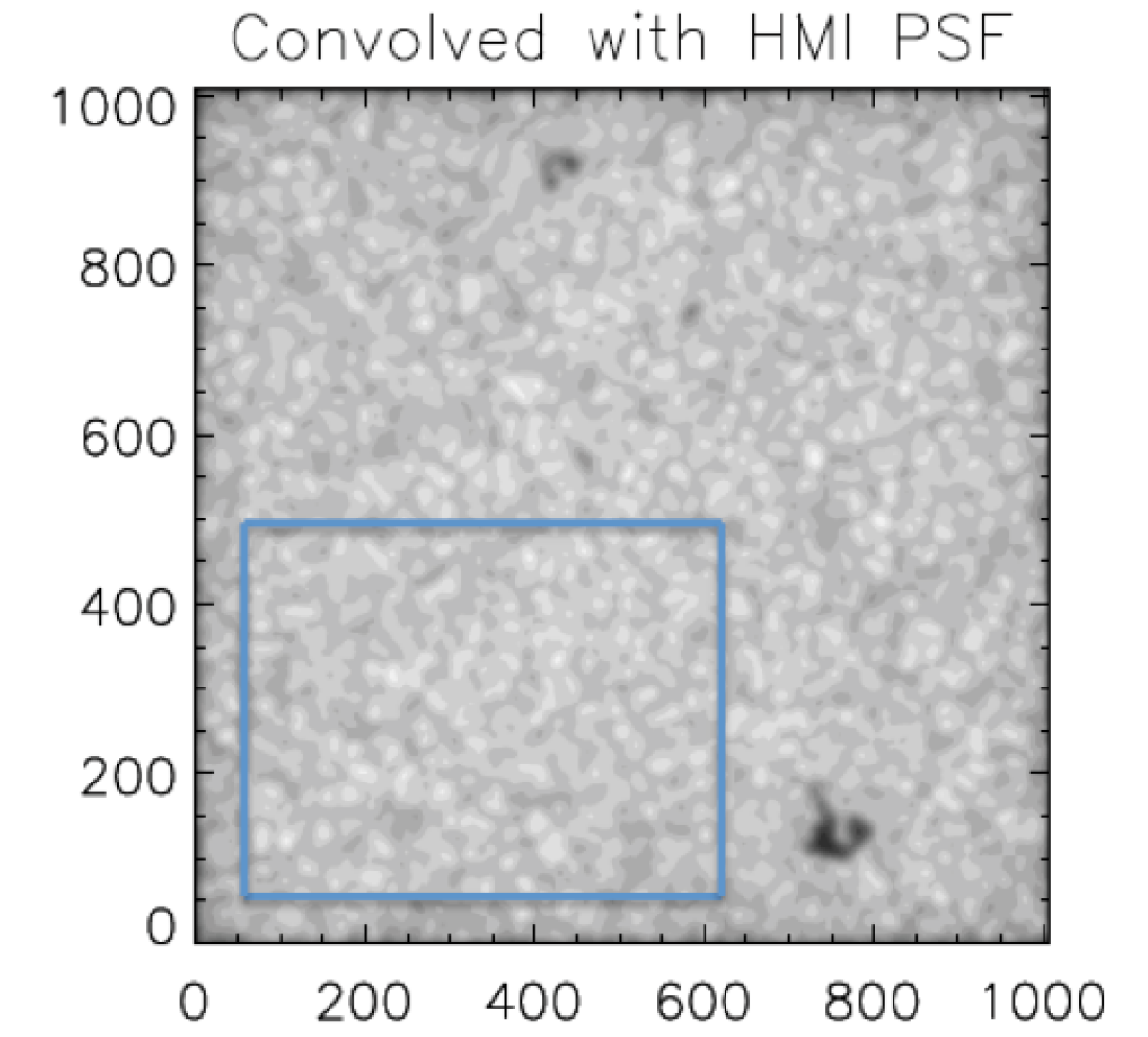}
             }
             \caption{A snapshot from the 6173 \AA~ line (spectral line synthesis was performed with the LTE LILIA code \citep{socas-navarro:2001}) from a numerical simulation of magneto-convection (courtesy of Bob Stein) for a 1008 $\times$ 1008 field of view with a grid size of 47.6 km per pixel. The panel at right shows the same data convolved with the HMI PSF. The granular contrast at left is 14.5\% while at right is 4.7\%. Note that HMI pixel size corresponds to about 370 km per pixel. The data on the right side are not rebinned since the convolution of the simulation data was performed with a PSF containing no information at higher frequencies than the HMI resolution scale. }
\label{fig:Fig6}
\end{figure}

The 45-second I$_c$ data also shows increased granulation contrast in the quiet-Sun, determined as the standard deviation divided by the average ($\sigma_{I_c}/ {\overline{I_c}}$). The contrast in granulation of the quiet Sun increased from 4.1 to 7.9\%, which is reported in Table~\ref{tab:table2}. 
We estimate the true solar photospheric granulation contrast by using a numerical simulation of the 6173 \AA~ Fe-I line as calculated by LILIA \citep{socas-navarro:2001} for a 1008 $\times$ 1008 field of view with a grid size of 47.6 km per pixel, see Figure~\ref{fig:Fig6} courtesy of Bob Stein  \citep{stein:2012}.  The granulation contrast of the simulation at this resolution was 14.5\%. Once the simulation data was convolved with the HMI PSF, a granulation contrast of 4.7\% was found, see Figure~\ref{fig:Fig6}. The simulated and observed values are not identical, with the observed granulation contrast (4.1\%, see Table~\ref{tab:table2}) being lower than the simulated values convolved with the PSF (4.7\%) and the HMI stray-light corrected data (7.9\%) being nearly half that of the simulated, high-resolution contrast (14.5\%). This is reassuring that the stray-light correction is, at least, not \textit{over-correcting} and falsely enhancing the contrast beyond what might be the true solar contrast which is unknown.  Note that granulation contrast varies as a function of center-to-limb angle. Standard deviations of $I_c$ for the 45-second and 720-second data near disk-center are reported in Table~\ref{tab:table2} and calculated as the values in a 200 $\times$ 200 pixel region at disk center when no active region is present.

\begin{figure}[ht]
  \centering
\includegraphics[trim= 10 10 0 0, clip, width=0.75\textwidth]{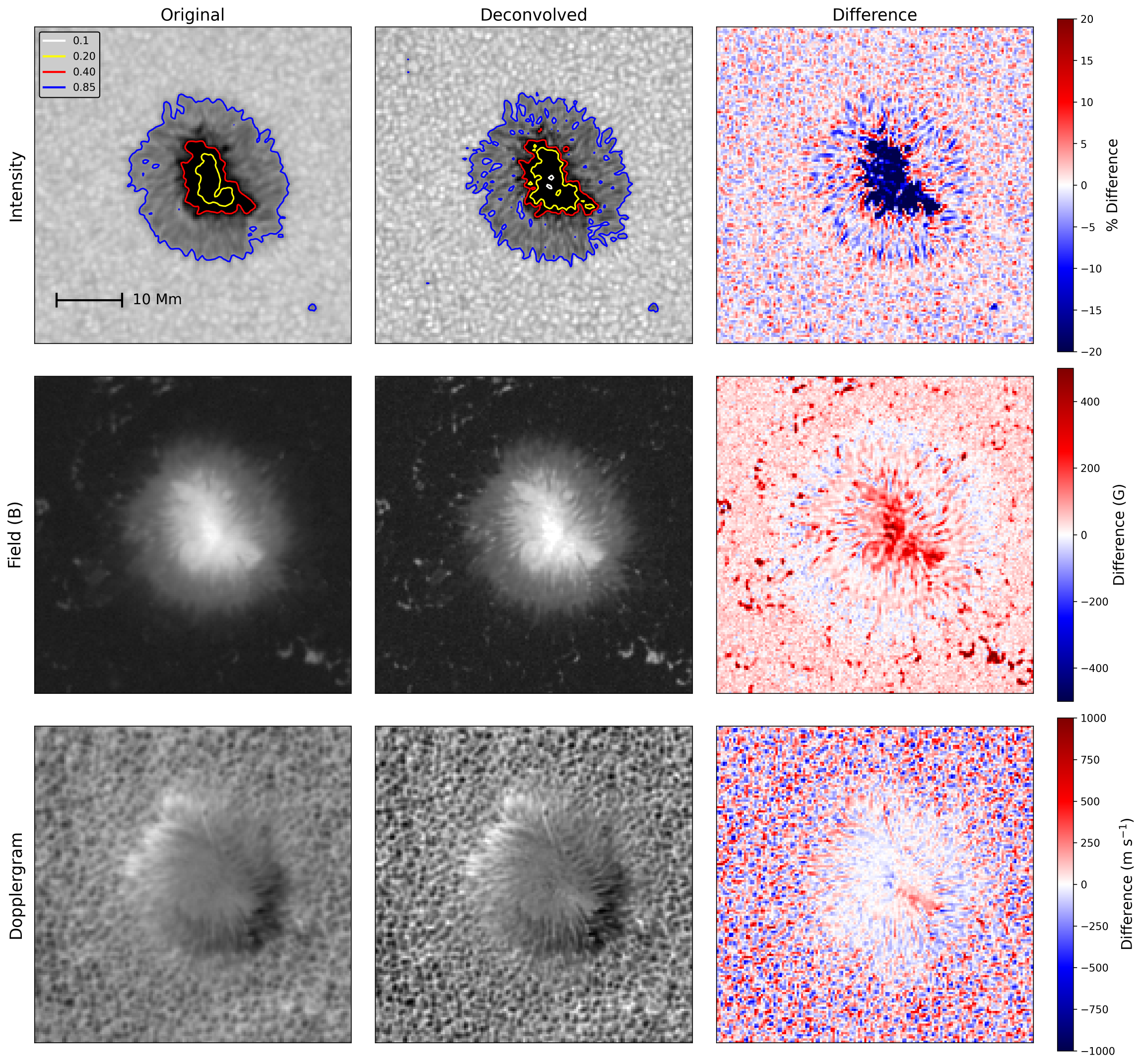}
\includegraphics[trim= 10 340 0 350, clip, width=0.75\textwidth]{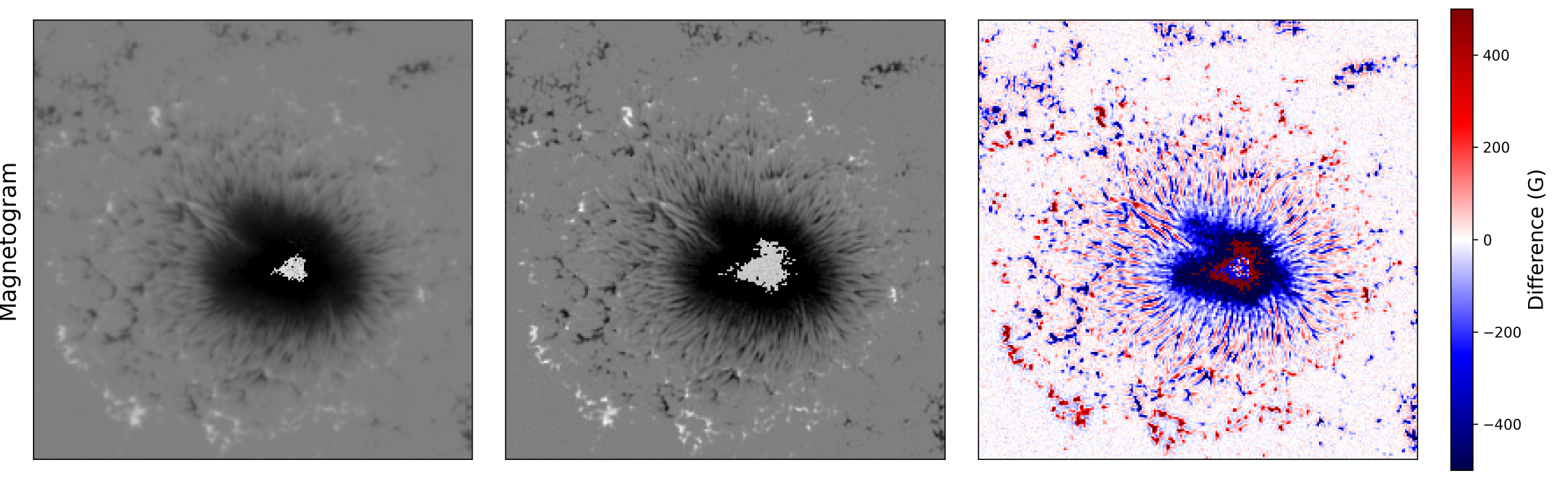}
\caption{Top 3 rows: A 150 $\times$ 150 pixel field of view  containing a sunspot (NOAA 13110) from 2022.09.28 19 UT is shown comparing the original (left), deconvolved (middle), and difference (right) for the 720-second $I_c$ data (top), field strength (middle) and Doppler velocity (bottom). Colored contours indicate the fraction of the quiet-Sun continuum intensity.  Differences are shown as a \% difference for $I_c$ (original minus the deconvolved divided by the original), in units of Gauss for B, and units of m s$^{-1}$ for the Dopplergram. Grayscale minimum and maximum values are the same for the left and middle plots. Bottom row: A 250 $\times$ 250 pixel field-of-view containing a sunspot (NOAA 11899) from 2013.11.18 17 UT is shown comparing the original (left), deconvolved (middle), and difference (right) for the 45-second line-of-sight magnetogram. Saturation, as seen in the umbra, is worse worse in the deconvolved data. Saturation occurs when a combination of the following conditions occur: $I_c$ levels fall to 10\% or less of the quiet-Sun continuum, the line depth is small, and the field strength is greater than 3000 Gauss.}
\label{fig:Fig7}
\end{figure}

Figure~\ref{fig:Fig7} shows a field of view that is 150 $\times$ 150 pixels  containing a sunspot (NOAA 13110) and surrounding quiet-Sun from 2022.09.28 19 UT. In the top row of Figure~\ref{fig:Fig7}, the original data (left), deconvolved data (middle), and percent difference (right) in original minus deconvolved data is shown for the 720-second $I_c$  data (top). Colored contours indicate the fraction of the quiet-Sun continuum intensity in the top row. Percent differences are shown as the original minus the deconvolved data values divided by the
original value for $I_c$. The differences in intensity are saturated in color at 20\% of the original values in Figure~\ref{fig:Fig7}. As expected, the deconvolved $I_c$ data show changes that enhance the contrast in granulation and in the penumbra. 

\begin{figure}[ht]
  \centering
  \includegraphics[trim=0 0 0 0, clip, width=.95\linewidth]{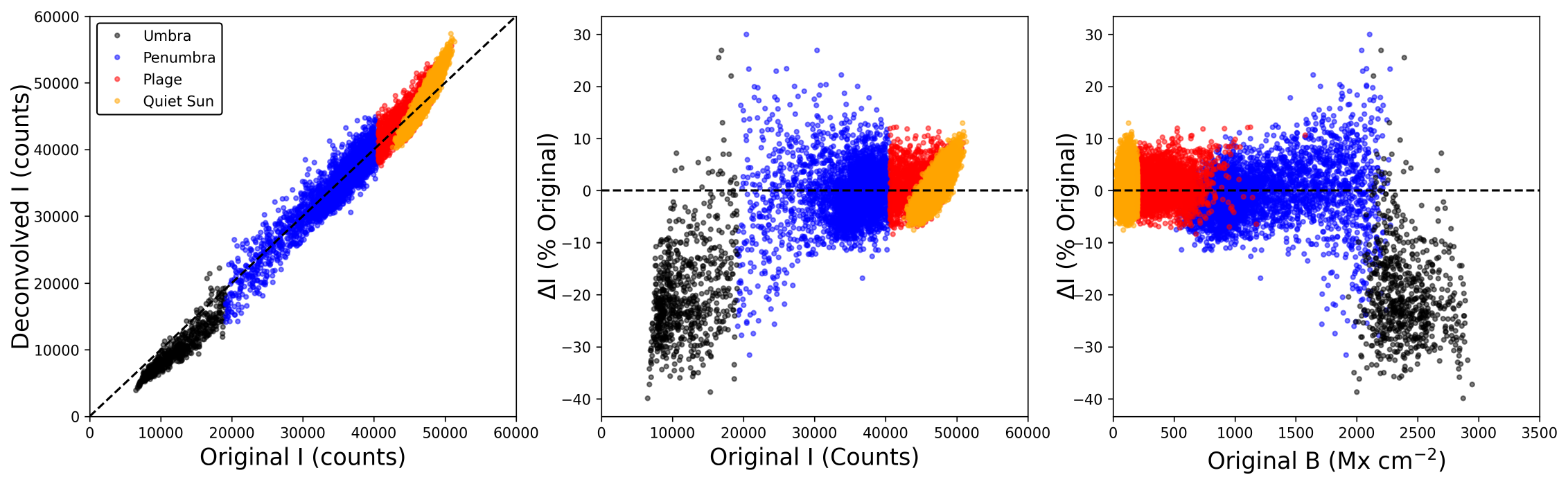} 
   \includegraphics[trim=0 0 0 0, clip, width=.95\linewidth]{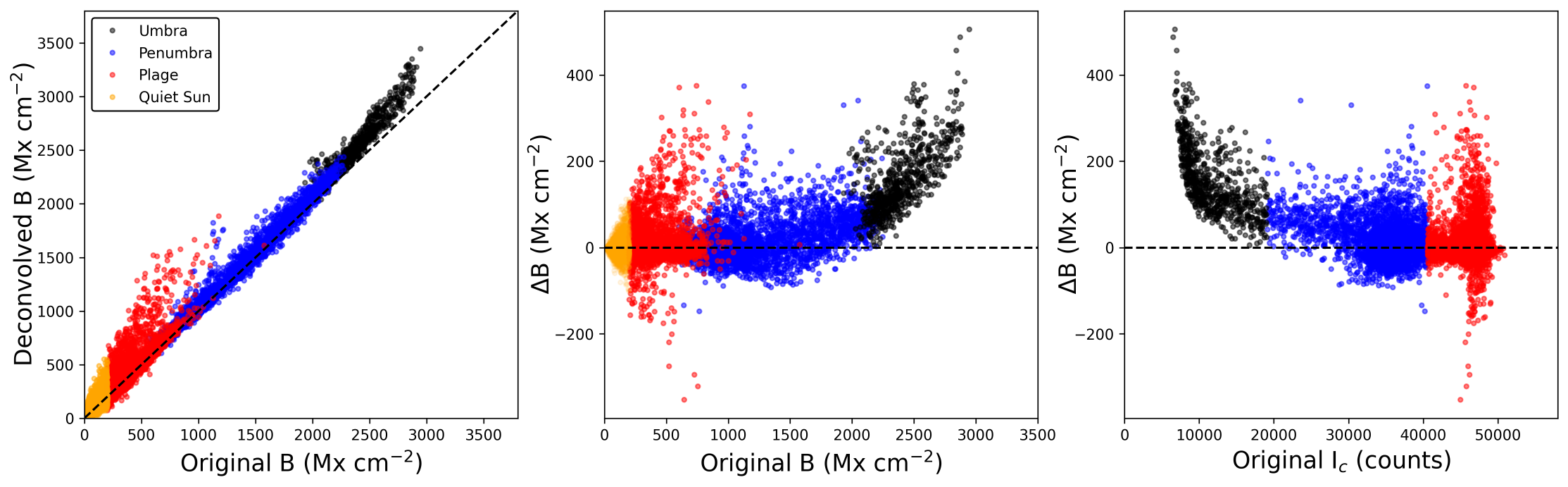} 
  \includegraphics[trim=0 0 0 0, clip, width=.95\linewidth]{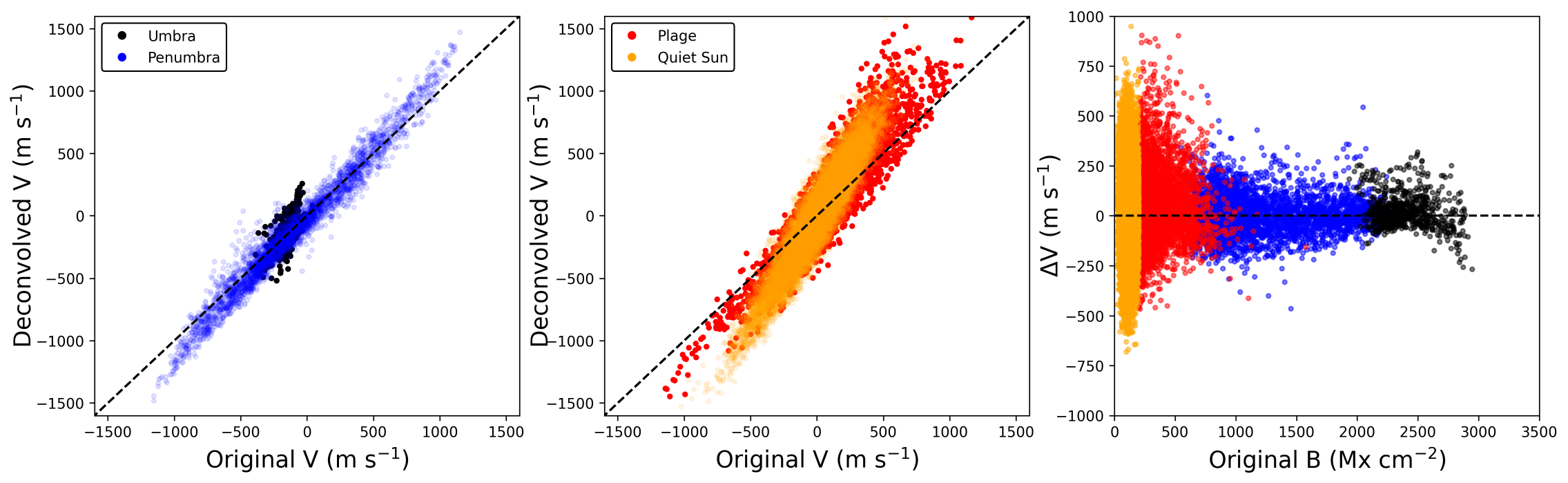}   
  \includegraphics[trim=0 0 0 0, clip, width=.65\linewidth]{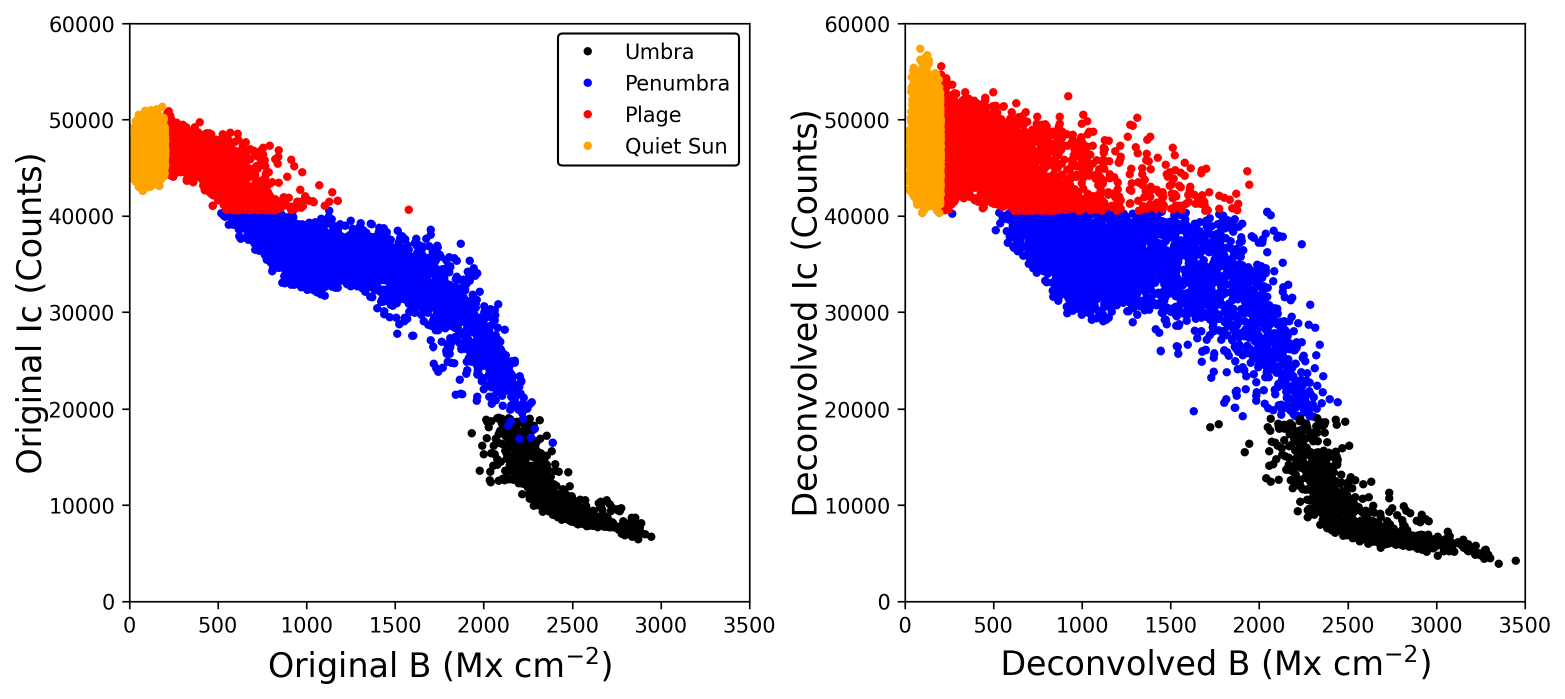}  
  \caption{Scatter plots explore the changes in continuum intensity (top row), magnetic field (second row), and Doppler (third row) after correction for scattered light for a 150 $\times$ 150 field of view containing an active region NOAA 13110 on 2022.09.28 at 19 UT. Values plotted on the x-axis (y-axis) are original (deconvolved) and show a variety of dependencies. The relationship between $I_c$ and B (lower row) is explored. The mean $I_c$ values remain essentially the same as these values are conserved but redistributed such that the dark features become darker and the bright features become brighter. Magnetic field values are not conserved in the same way with field strengths increasing after correction. }
  \label{fig:Fig8}
\end{figure}

Figure~\ref{fig:Fig8} shows scatter plots of the NOAA 13110 data shown as images in the top panels of Figure~\ref{fig:Fig7}. The top row in Figure~\ref{fig:Fig8} illustrates how the umbra only decreases in $I_c$, while the penumbral filaments both darken and lighten depending on their original brightness relative to the average brightness and the granular contrast also increases. 

The light levels in the dark core of the umbra decreased from 5.5\% of the nearby quiet-Sun continuum intensity to 3.3\% after deconvolution. The inferred minimum temperature in the umbra associated with that intensity decrease is a change of 230 K, from 3370 to 3140 K accordingly. 

In the $I_c$ versus B scatter plot in Figure~\ref{fig:Fig8}, bottom row, there is a wider range of $I_c$ values on the y-axis after deconvolution compared to the original range.  The mean $I_c$ values of the different features are the same in the original and deconvolved data but there is wider distribution of values. 


\FloatBarrier
\subsubsection{Limb Fit}

As a routine part of the data processing, the image center coordinates are determined and a fit to the solar limb is performed to find a value of the solar radius. To understand the limb fit, a description of a few keywords and their determination is in order. To find the image center coordinates, a two-dimensional cross correlation is performed between the image and itself rotated 180$^{\circ}$. The location of the peak cross correlation amplitude is found,  converted into the image center coordinates, and ultimately recorded as the CRPIX1 and CRPIX2 keywords. See Appendix for a discussion of keywords associated with the limb-fitting procedure. 

We report small changes in the values calculated for the deconvolved data in the limb fitting process. The radius of the Sun does not change in a significant way before and after deconvolution. The image center values change on the order of 0.01 pixels (standard deviation) which is well within the acceptable margin of error for image co-registration of 0.05 pixels. While the shift in the image center in the y-direction is evenly distributed around zero, there is a systematic shift in the x-direction of the image center towards the left side of the instrument, or east limb when in normal pointing mode, indicating that there is an asymmetry of the light levels in the instrument, as already mentioned in the off-center contour levels in the disk of Venus shown in Figure~\ref{fig:Fig3}. This motivates future work in which the the PSF is developed with an azimuthal dependency. 

\subsection{Magnetic Field Data Products}\label{sec:sec4.2}

Line-of-sight magnetograms as well as vector magnetic field data products were analyzed to determine the changes due to stray-light correction.  Unlike the $I_c$ data, the corrected magnetic values do not retain the same mean as the original data, but increase as the light scattered from the brighter, non-magnetic or lower magnetic field strength area is removed from the umbrae, penumbrae and plage. Figure~\ref{fig:Fig7}, second row, shows the total field strength before and after deconvolution. Note that the difference (second row, right panel) is dominated by increased field values. The standard deviations of the deconvolved data are twice as high the standard deviations of the original data, see Table~\ref{tab:table2}. 
\begin{figure}[b]
  \centerline{\includegraphics[trim=00 10 00 00, clip, width=.95\linewidth]{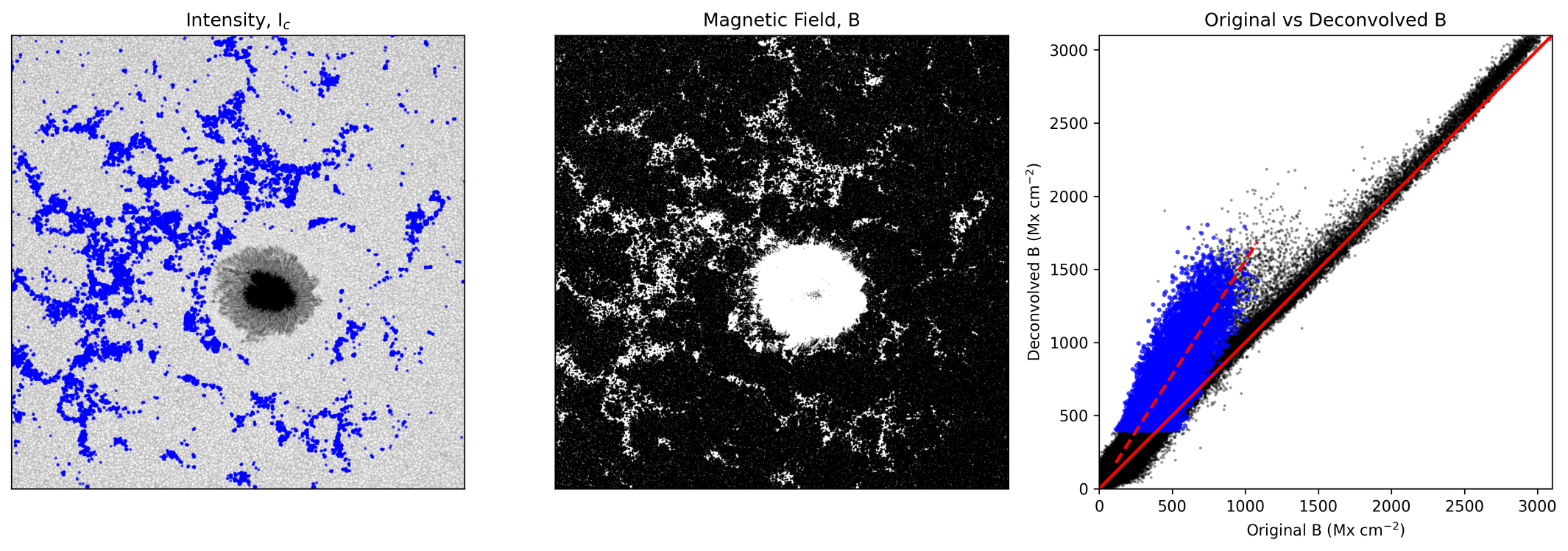}}
\caption{Location of plage are denoted as blue points in a 700 $\times$ 700 pixel I$_c$ image (left) with a corresponding field strength image, saturated at a low value of 400 Mx cm$^{-2}$, to show location of plage (middle). A scatter plot of VFISV solutions for the total field strength, B, of original and deconvolved data is shown (right). The slope of the red dashed line, $m=1.56$, represents a fit to the deconvolved and original plage field strengths. Plage are defined as locations where the deconvolved field strength is greater than or equal to 400 Mx cm$^{-2}$, the intensity is greater than 0.9 of the mean intensity, and it is located at least 100 pixels away from the center of the umbra (otherwise penumbral fields will be included). The field strengths for penumbra and umbra are less affected than those of plage, as evident in the field strengths in the range of 2000-3000 seen in the scatter plot with a solid red line showing unity.  }
\label{fig:Fig9-plage}
\end{figure}
Scatter plots associated with the NOAA 13110 images in Figure~\ref{fig:Fig7} are shown in Figure~\ref{fig:Fig8} for the original versus deconvolved $B$ in the second row, left column. Plage, penumbral and umbral values increase up to 450 Mx cm$^{-2}$, see the second row, middle $\Delta$B panel.  Another common way to view the $B$ values in and around a sunspot are as a function of $I_c$, as seen in the lower row of Figure~\ref{fig:Fig8}, with the original (left) and deconvolved (right) dependency shown. The umbra darkens and its field values increase. The penumbra, plage and quiet-Sun show an increased range in $I_c$, maintaining the mean, with stronger field strengths. 

The largest changes were found in the values of plage field strength from the vector data as found in the hmi.B\_720s\_dconS data. An example is shown in Figure~\ref{fig:Fig9-plage} that shows a 600$\times$600 pixel region from 2013 November 18.  The plage locations are indicated in the $I_c$ image (left) as blue dots and the resultant change in the B data is shown for the entire field-of-view with the plage values increasing $\sim1.5$ times the original value. The plage are more affected by scattered light as their locations are nearer to bright, quiet-Sun pixels. Meaning, plage pixels are often located within a steep, spatial gradient in intensity compared to pixels in umbra, for example, where the gradient is less steep from pixel to pixel.

The algorithm used in MDI and HMI for determining the line-of-sight magnetogram values can return anomalous values in sunspot umbrae where the intensity counts are low. The line becomes very shallow and the Zeeman splitting of the spectral line is large due to a strong magnetic field \citep{Liu:2007}.  This effect is sometimes termed magnetogram `saturation'; see NOAA 11899 images in the bottom row in Figure~\ref{fig:Fig7} which shows a 250 $\times$ 250 pixel field of view of a large, strong sunspot surrounded by plage. This sunspot is particularly affected by saturation. While the majority of sunspots observed by HMI are not negatively affected by saturation, any very dark sunspot with a field strength significantly above 3000 Mx cm$^{-2}$ can have saturation. The deconvolution removes light from the umbrae which increases the number of pixels affected by saturation, see the increased number of pixels affected in the deconvolved data (middle column of NOAA 11899 data) of Figure~\ref{fig:Fig7} compared to the original data (left column).  

%The trends in the corrected $I_c$, $B_{los}$ and $V$ values are similar to what is seen in NOAA 13110 also shown in Figure~\ref{fig:Fig7} but note that for NOAA 13110,   the $B$ values are shown while for NOAA 11899, the $B_{los}$ values are shown. 

\begin{figure}[htbp]
  \centerline{\hspace*{0.015\textwidth} 
              \includegraphics[trim= 0 3 0 0 , clip, angle=0,width=0.65\textwidth]{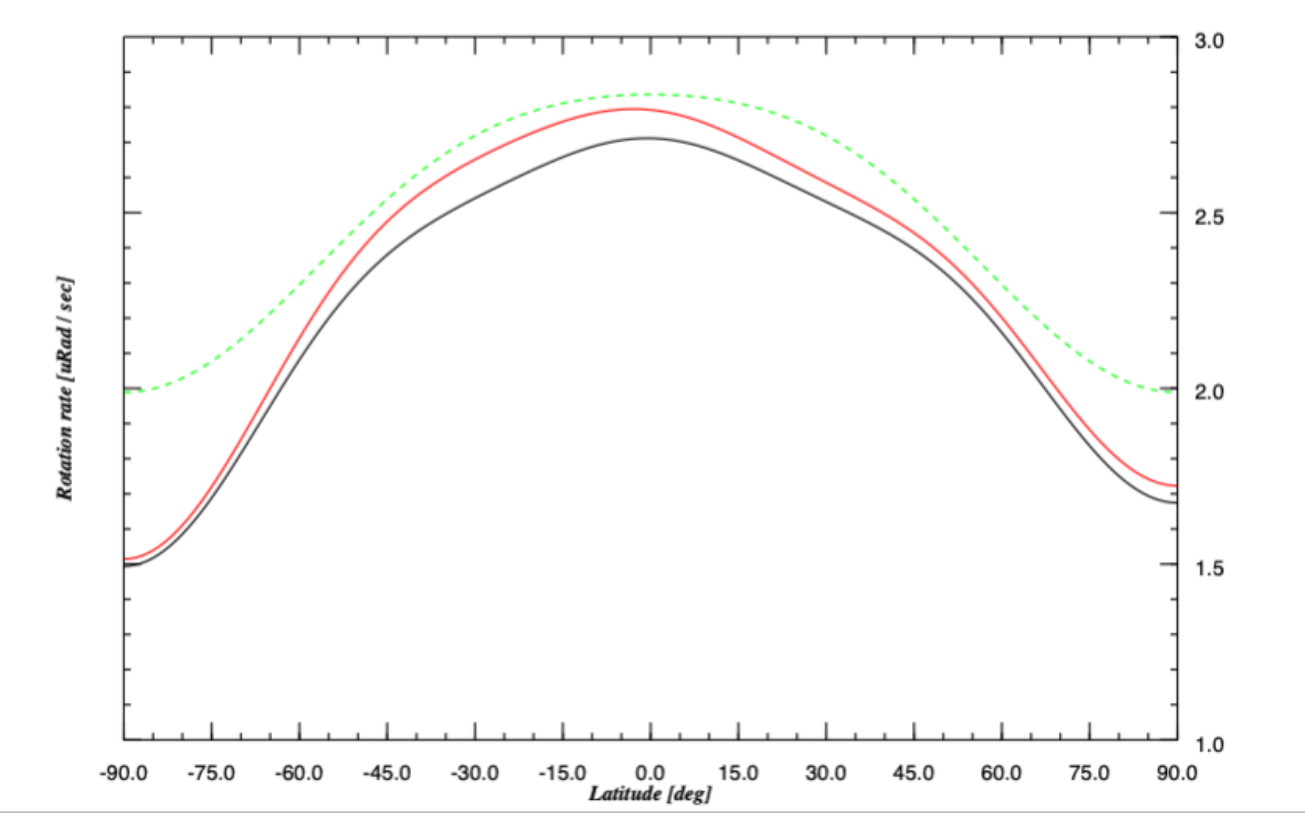}
               }
\caption{A fit of rotation coefficients to the mean Doppler signal from a 1/3-rotation average of deconvolved data centered at Carrington time 2266:360 (2023.01.01 09:04 UT) results in a significantly faster equatorial rotation rate (red line) and steeper and a slightly more asymmetric differential rotation profile than the fit to the original data (black line). Both fits are to 6th order in Chebyshev polynomials of $\sin$(Latitude). The green dashed curve is the commonly used ``Snodgrass" profile \citep{snodgrass:1984}. }
\label{fig:Fig10}
\end{figure}

\subsection{Doppler Velocities}\label{sec:sec4.3}

After deconvolution, the quiet-Sun Doppler velocities show better resolution of the downflows in the inter-granular lanes. The phenomenon in which the brighter, hotter, upwelling material dominates the signal in quiet-Sun in the uncorrected data is known as the convective blueshift. The deconvolution partially corrects this, as shown in the Doppler plots of Figure~\ref{fig:Fig7}. The changes in the amplitudes are up to $\pm$1 km s$^{-1}$ for the correction, see Figure~\ref{fig:Fig8} (third row, middle column). The standard deviation values for the original and deconvolved data, for the 45-second and 720-second observables, are found in Table~\ref{tab:table2}.

 As a sanity check that the scattered-light correction does not spuriously alter the velocity data, we analyzed a $1024\times1024$ quiet-Sun field-of-view near disk center, using one original and one deconvolved image per day throughout 2024. For each image we computed the mean intensity at the six wavelength tunings, after masking out pixels with $|B|>200$ G. Across the year, the differences (original minus deconvolved) in mean intensity at each tuning are $0.00$–$0.28\%$ (typically $<100$ counts out of $\sim45{,}000$). The averaged line profile shows no measurable change in shape or in line center within the measurement uncertainties.

The local helioseismology analysis (discussed in Section \ref{sec:sec4.5}) involves detrending the Doppler data by a long-term average at each image location in order to remove the first-order effects of solar rotation and image-dependent systematics as the target regions are tracked. We produced a long-term average of solar rotation over one-third of a solar rotation (about 9 days) centered at a target Carrington time (2023.01.01 09:04 UT) from the deconvolved data. Comparing this average rotation rate with that for the original data, it was noticed that there was a systematic difference of order 50 $m s^{-1}$ over most of the disk, declining from about 60 $m s^{-1}$ at disk center to about 15 $m s^{-1}$ at a center-to-limb distance of 0.95, in the sense of the deconvolved data exhibiting a larger apparent redshift over most of the disc. There was also a small but significant east-west variation, so that at the east limb the deconvolved data actually exhibits a slight blue shift of a few $m s^{-1}$ relative to the unconvolved data. This means that any fit of the Doppler data to the surface differential rotation that must take into account the limb shift will produce a slightly different curve. Such an effect is illustrated in Figure~\ref{fig:Fig10}.

\begin{figure}[ht]
  \centering
  \includegraphics[trim=70 0 98 50, clip, width=.25\linewidth]{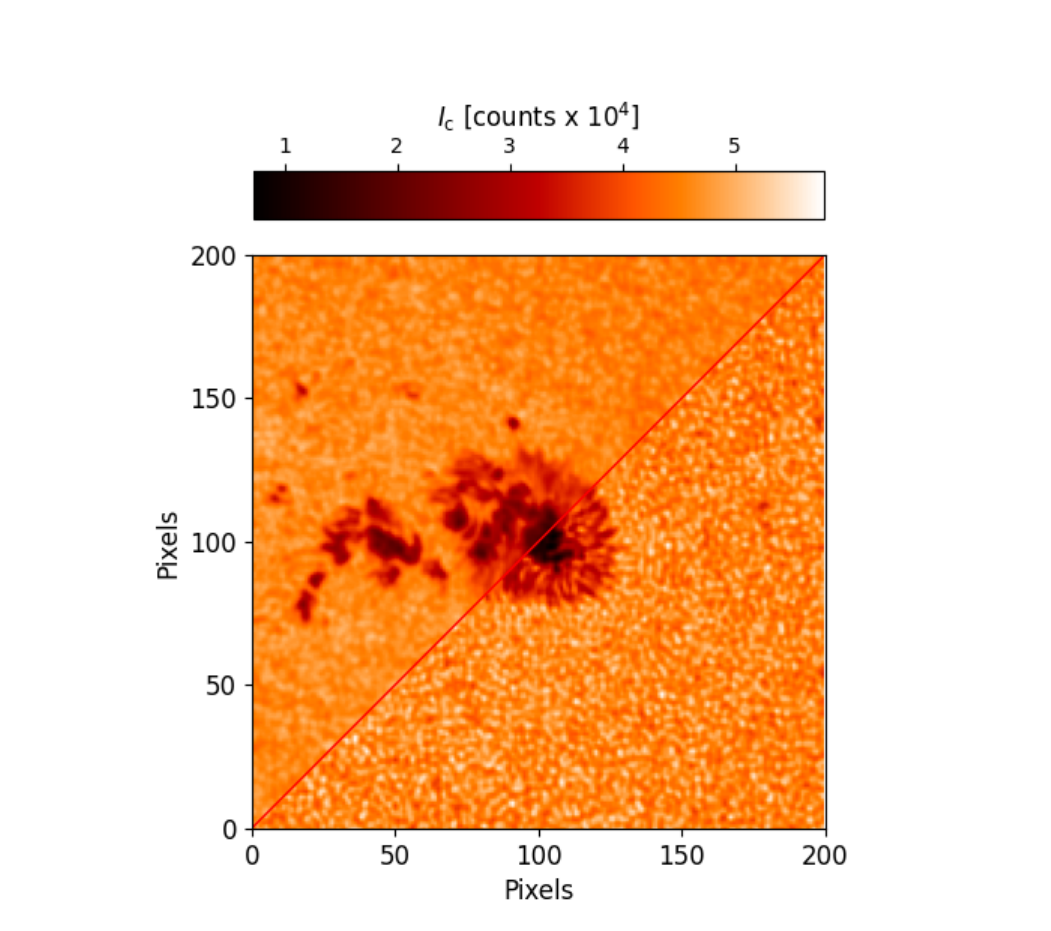} \qquad
  \includegraphics[trim=70 0 98 50, clip, width=.25\linewidth]{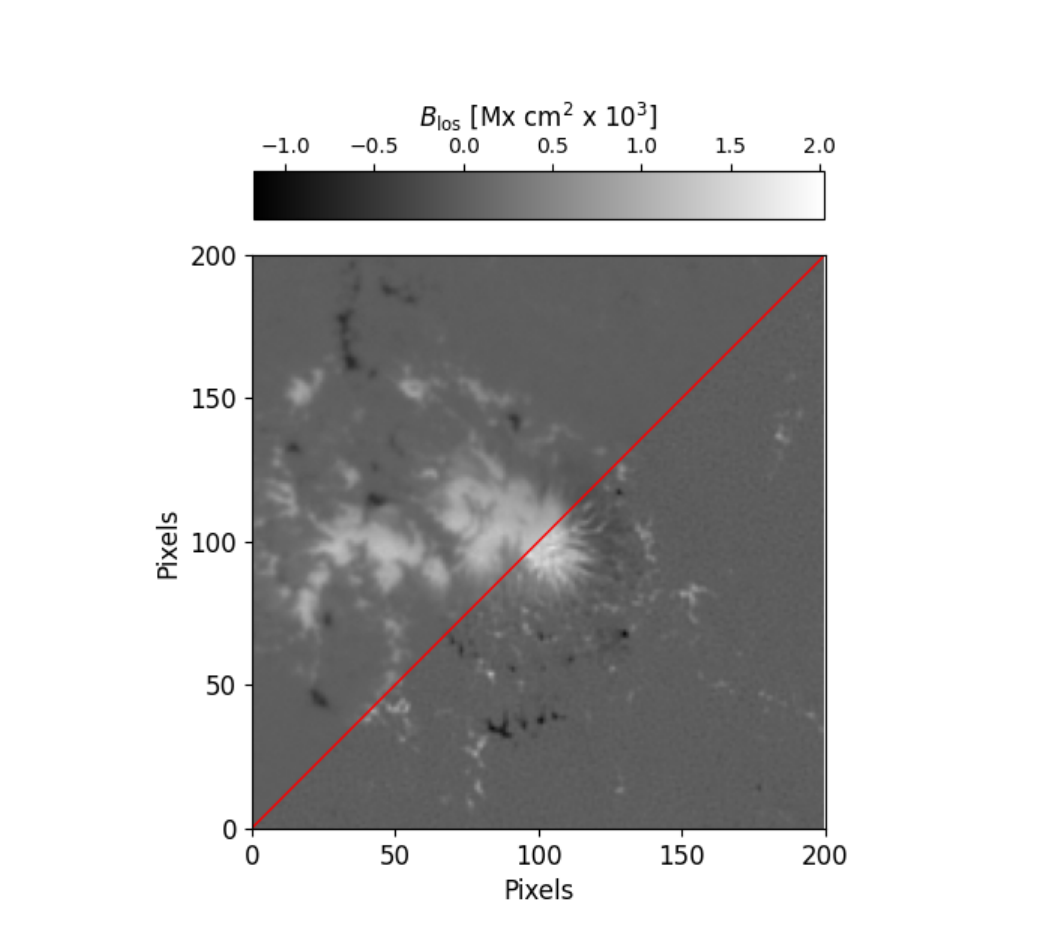} \qquad
  \includegraphics[trim=70 0 98 50, clip, width=.25\linewidth]{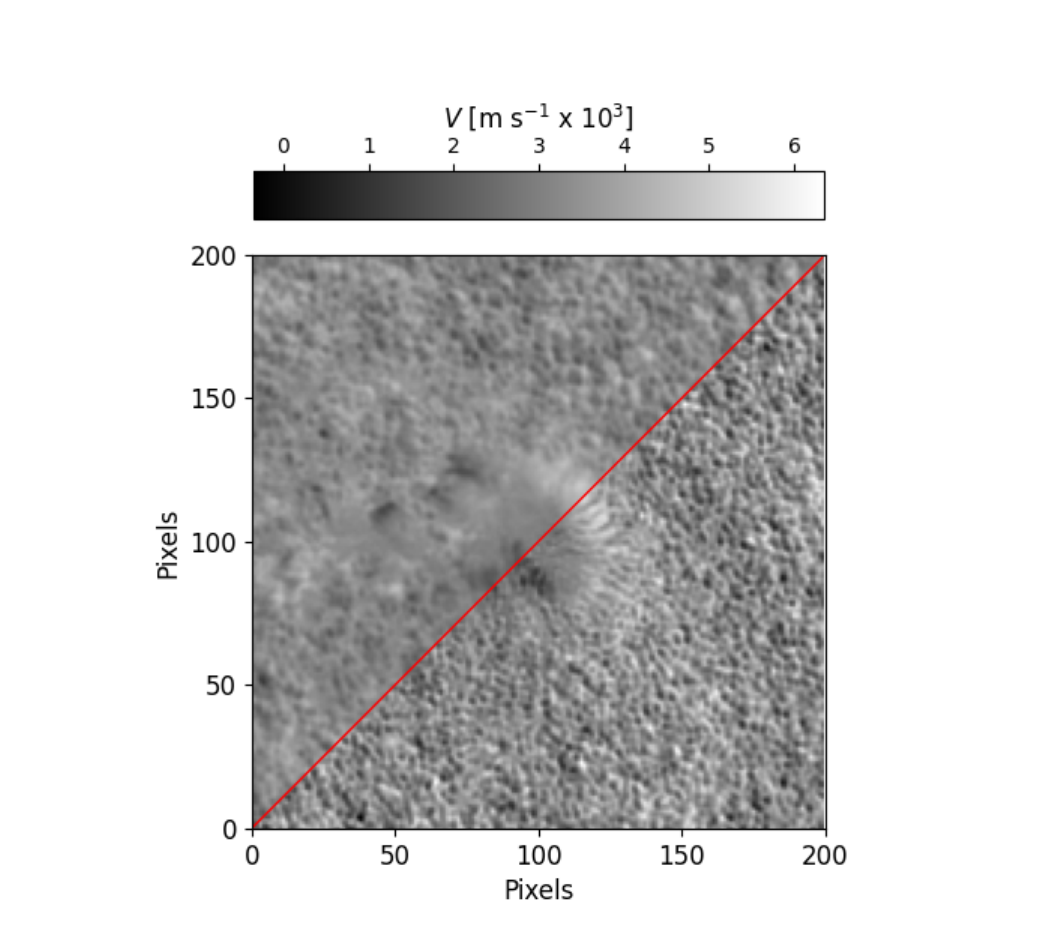} \qquad
  \includegraphics[trim=30 0 68 40, clip, width=.25\linewidth]{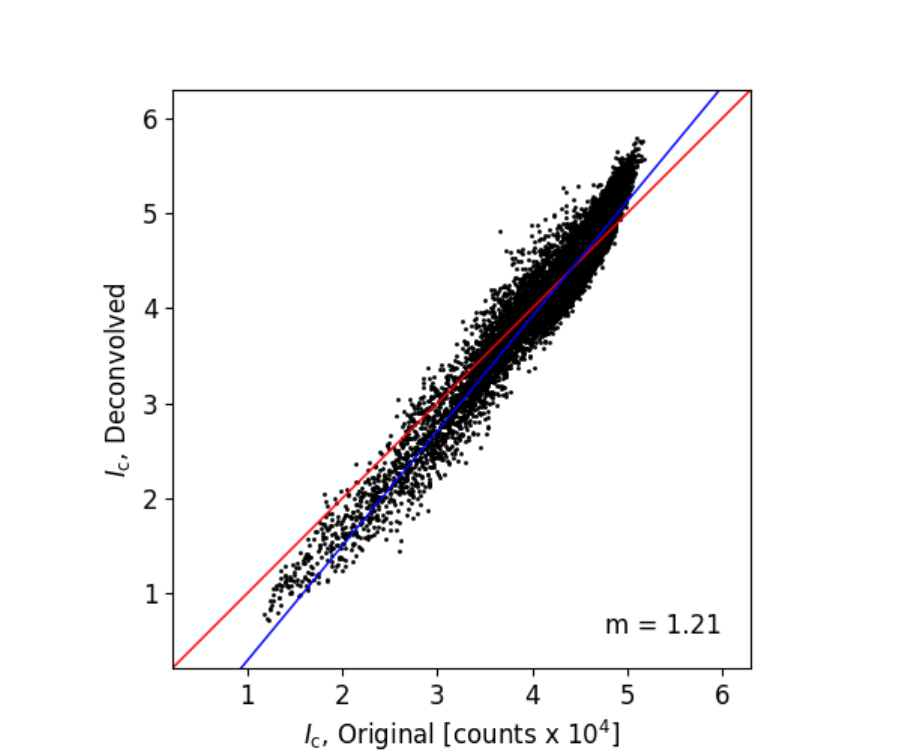} \qquad
  \includegraphics[trim=30 0 68 40, clip, width=.25\linewidth]{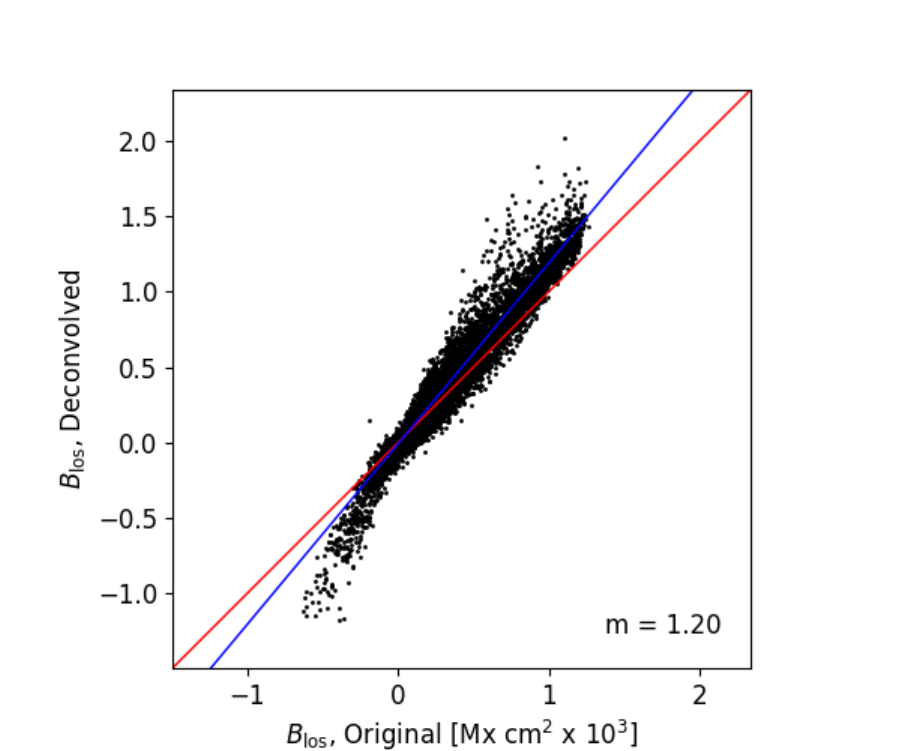} \qquad
  \includegraphics[trim=30 0 68 40, clip, width=.25\linewidth]{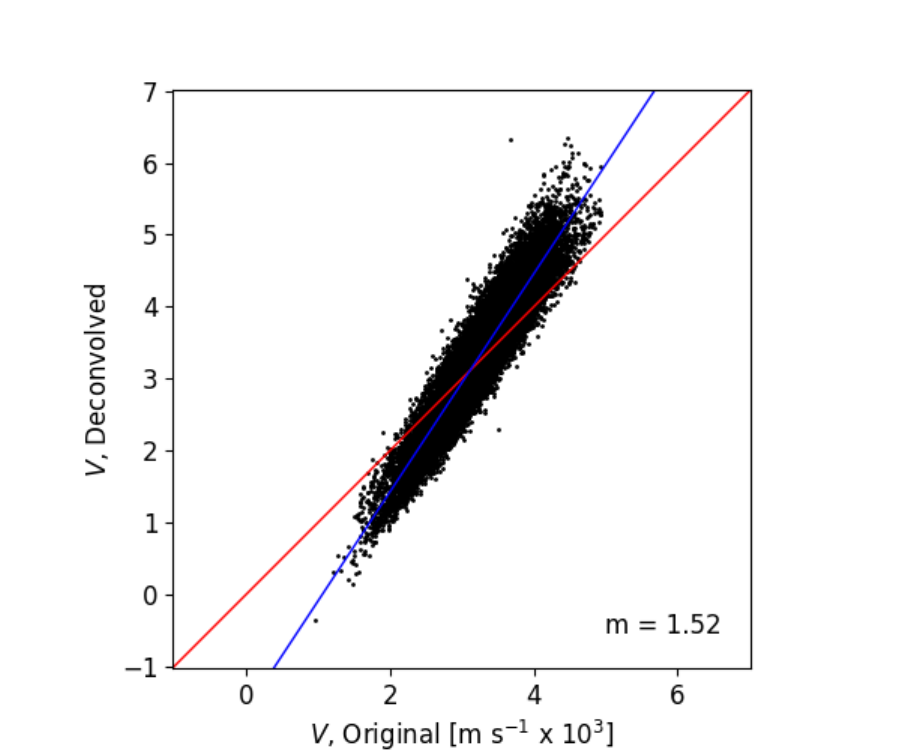} \qquad
  \rule{\linewidth}{0.5pt}
   \par\vspace{0.5em}  % add vertical space after line
  \includegraphics[trim=70 0 98 50, clip, width=.25\linewidth]{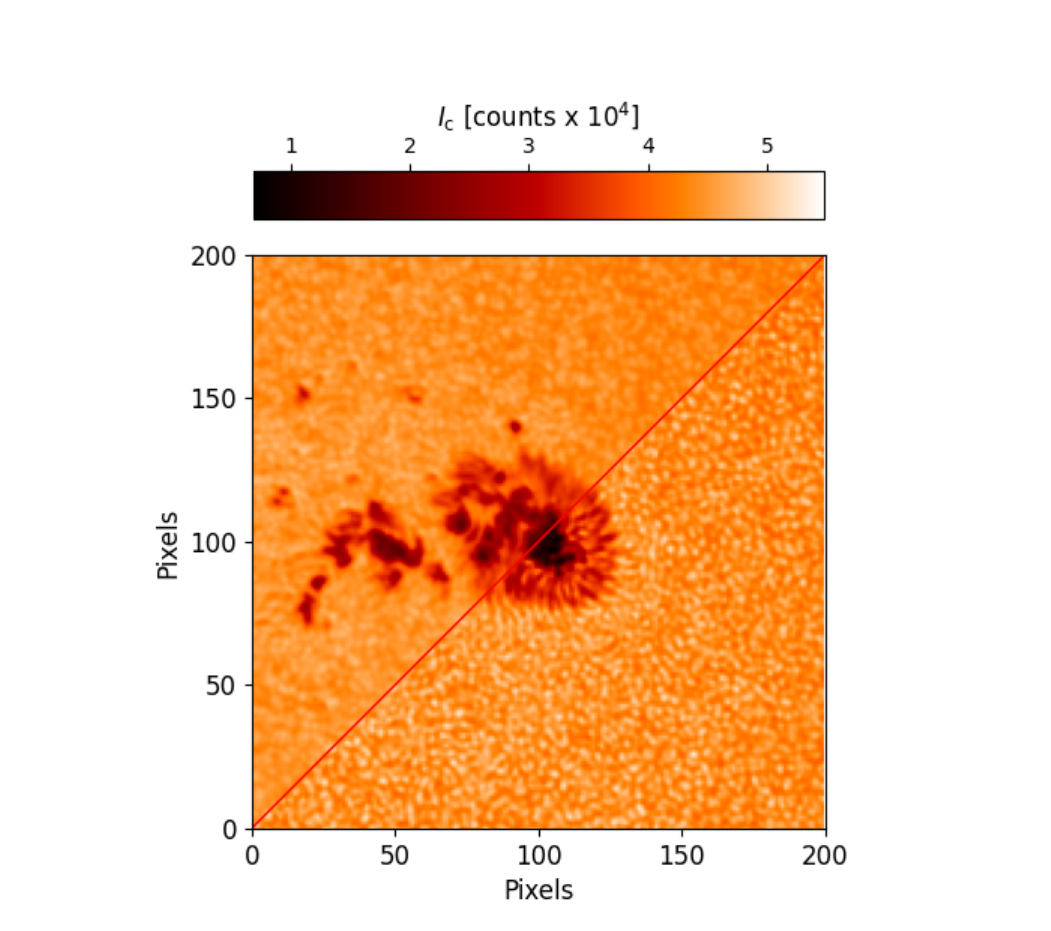} \qquad
  \includegraphics[trim=70 0 98 50, clip, width=.25\linewidth]{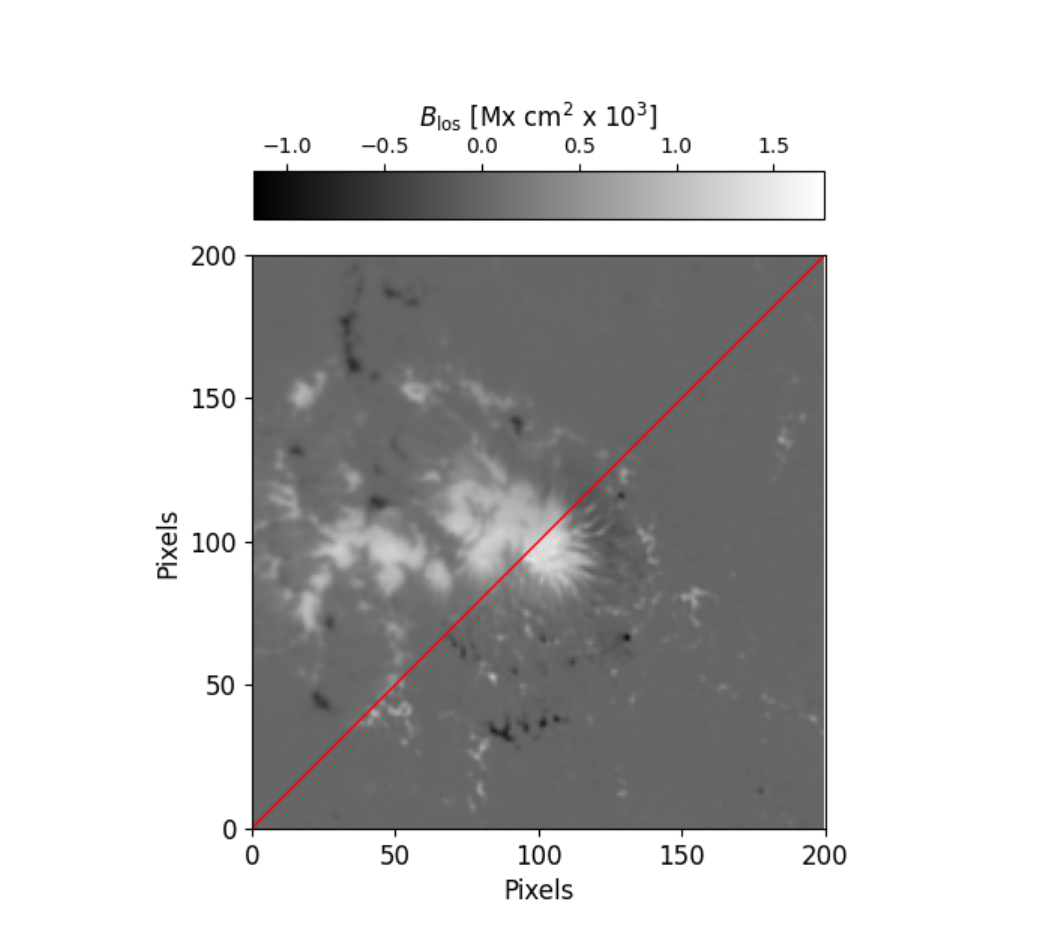} \qquad
  \includegraphics[trim=70 0 98 50, clip, width=.25\linewidth]{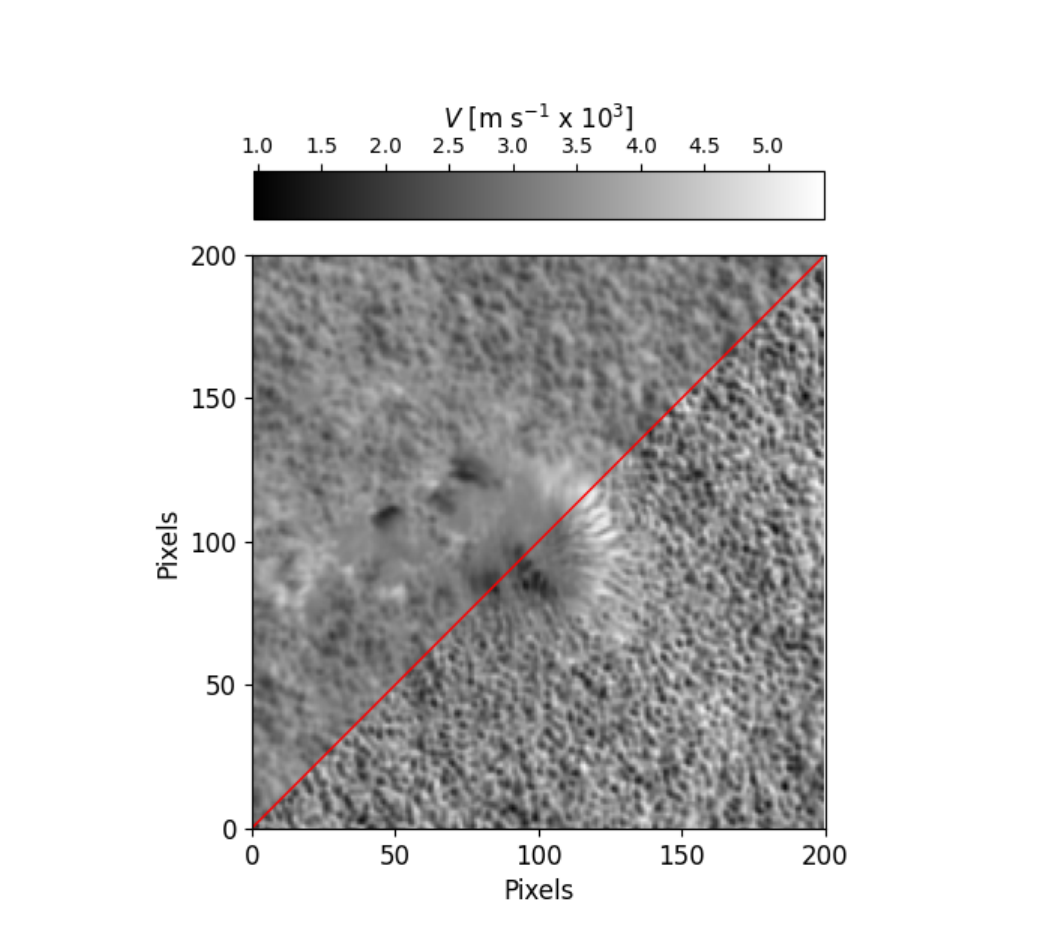} \qquad
  \includegraphics[trim=30 0 68 40, clip, width=.25\linewidth]{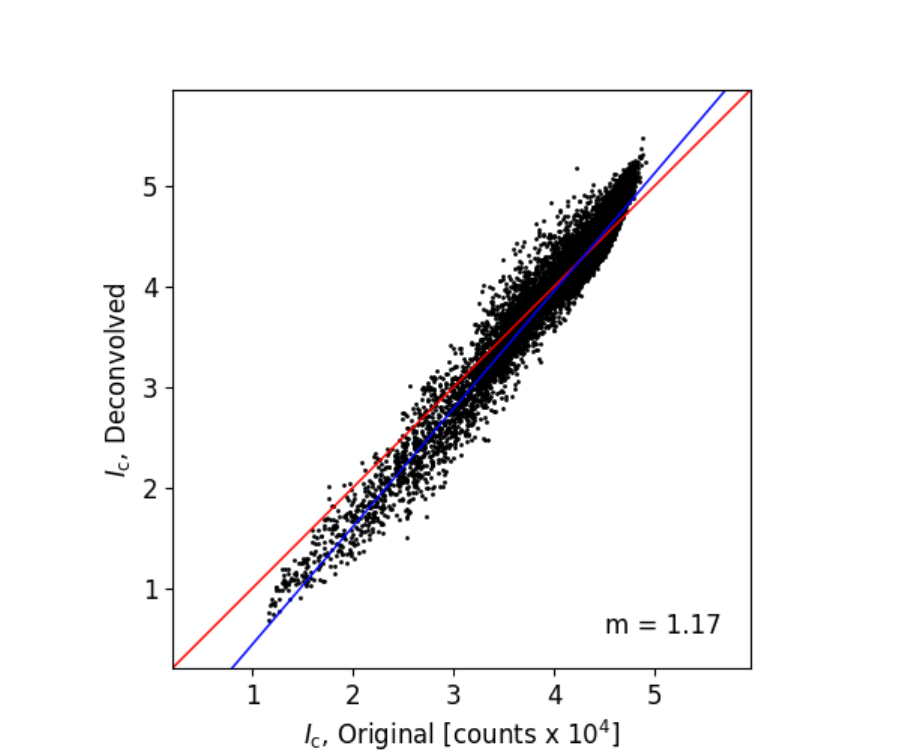} \qquad
  \includegraphics[trim=30 0 68 40, clip, width=.25\linewidth]{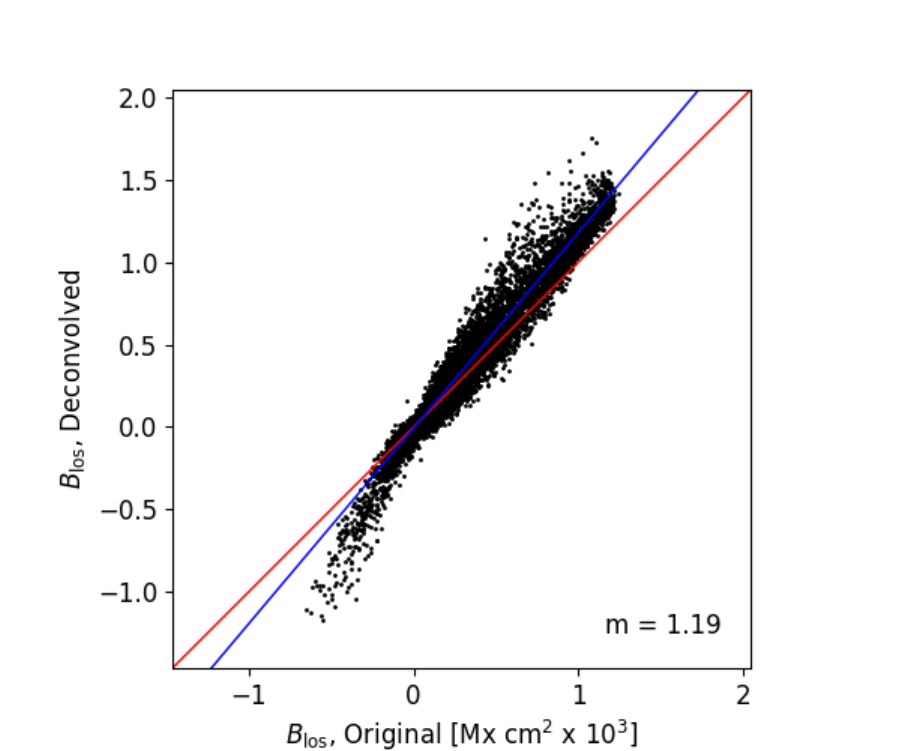} \qquad
  \includegraphics[trim=30 0 68 40, clip, width=.25\linewidth]{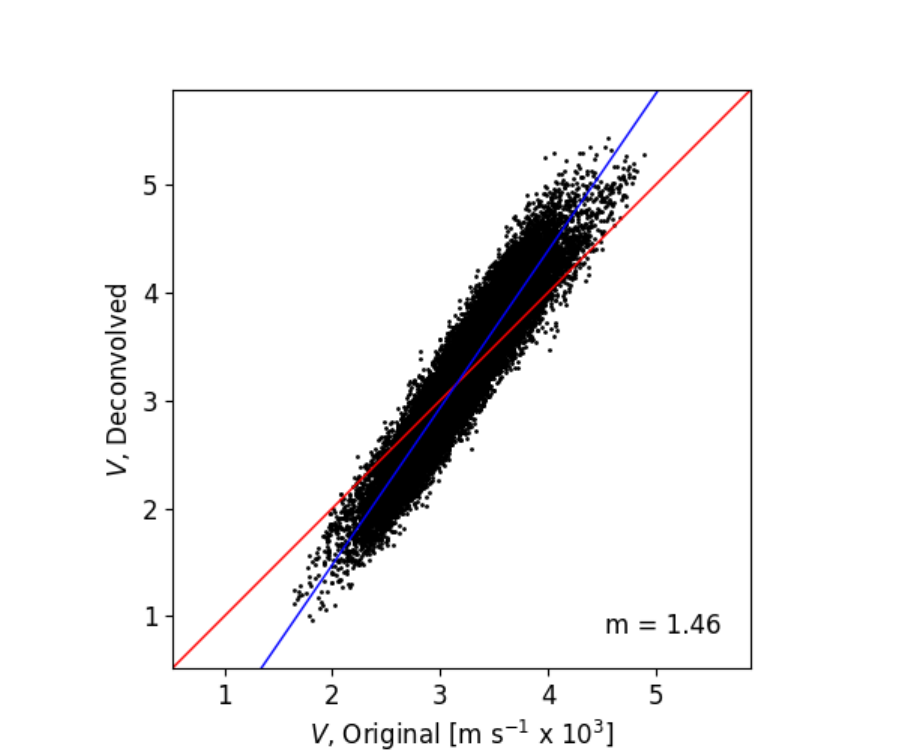} \qquad
  \caption{Comparison of the 45-second cadence data (upper two rows that are above the horizontal line) with the 720-second data (lower two rows) with Ic, Blos, and V shown with the original data (upper, left) and deconvolved data (lower, right) in each frame. Scatter plots are shown of the original versus deconvolved values for the 200 $\times$ 200 pixel region shown in the row just above it. Doppler velocities do not have the average solar rotation value subtracted so are positive due to the field-of-view position on the solar disk. The slope corresponding to a linear fit shown in blue is provided with slope, $m$, shown within the panel. The 45-second and 720-second data show very similar trends after the correction. The 720-second data has a slightly lower slope for all quantities, probably due to the 720-second data being a longer time average which filters out the contributions from the 5-minute $p$-modes.}
  \label{fig:Fig11}
\end{figure}

\begin{table*}[hb]
\centering
\begin{threeparttable}
\caption{Standard Deviations of Data}
\begin{tabular}{lllll}
\hline 
\multicolumn{1}{l}{\textbf{Observable}} &
\multicolumn{1}{l}{\textbf{720s Orig}} &
\multicolumn{1}{l}{\textbf{720s dconS}} & 
\multicolumn{1}{l}{\textbf{45s Orig}} & 
\multicolumn{1}{l}{\textbf{45s dcon}}  \\
\hline
$\sigma(I_c)$ & 1183 & 2280 & 2806 & 5471  \\
$\sigma(I_c)$/$I_c$ & 2.6\% & 4.9\% & 4.1\% & 7.9\%  \\
$\sigma(B_r)$ &28.7 & 48.6 & NA & NA \\
$\sigma(M)$ & 14.3 & 25.8 & 16.2 & 27.1 \\
$\sigma(V)$ & 199 & 395 &389 & 671\\
%Br offset & 1.83 & 1.73 & - & - & 0.3 & 0.2 \\
\hline 
\end{tabular}
\begin{tablenotes}
\footnotesize
\centering
\item ~~~~~~~~~Values determined from a quiet-Sun region of 200$\times$200 pixels at disk center. 
\end{tablenotes}
\end{threeparttable}
\label{tab:table2}
\end{table*}
%\FloatBarrier

\subsection{Comparison of 45-second and 720-second Data}\label{sec:sec4.4}

As discussed in Section~\ref{sec:sec3}, the 45-second and 720-second data are corrected in different ways. The \_dcon data have the correction applied to individual filtergrams for the 45-second data and the \_dconS corrections are applied to the average Stokes profiles for the 720-second data, see Figure~\ref{fig:Fig11} for a comparison of the 45-second and 720-second data. They compare favorably with some expected deviation due to the longer averaging for the 720-second data and in particular, that the 720-second data filters out contributions from the 5-minute $p$-modes. Tests were also performed to compare 720-second data derived from deconvolution applied to every filtergram versus deconvolution applied to the Stokes profiles and the results were identical, albeit the expected noise due to the inversion process. 

\subsection{Local Helioseismology}\label{sec:sec4.5}
 %\subsection{Ring Diagrams}
 \label{sec_RD}

\begin{figure}[ht]
\centering 
\includegraphics[width=0.80\textwidth]{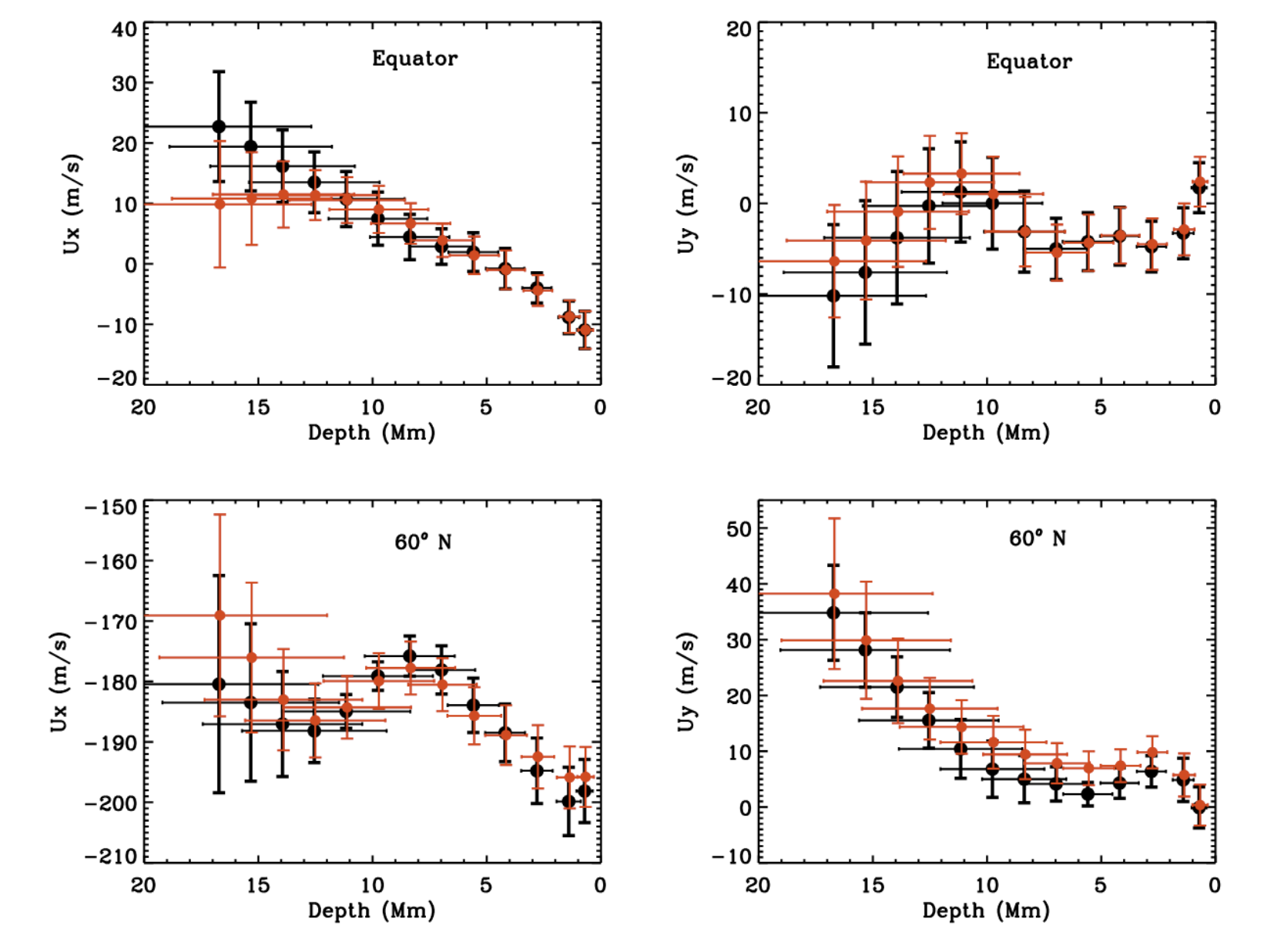}
\caption{
Comparison of zonal (left panels) and meridional (right panels) components of subsurface velocity, relative to a Carrington-rotating frame, as determined from ring-diagram analysis. Flow inversions are derived from Doppler data, averaged over six days around January 1, 2023, within circular regions of 15$^\circ$ heliographic diameter. The top panels show results for a region centered at the central meridian and equator. The bottom panels display results for a region centered at the central meridian and at 60$^\circ$ N latitude. Black and red data points represent flows as a function of depth from original and deconvolved data, respectively. Error bars indicate the error of the mean.
}
\label{fig:Fig12}
\end{figure}

Scattered-light corrected Doppler velocity data from 2022 December 28 to 2023 January 4 were processed using both the HMI time-distance helioseismology pipeline \citep{Zhao:2012SP} and the HMI ring-diagram analysis pipeline \citep{Bogart:2011}. The resulting zonal and meridional velocity measurements from these deconvolved data were then compared directly with the results from the original HMI Dopplergrams. Figure~\ref{fig:Fig12} shows the results obtained by ring-diagram analysis using regions of diameter 15$^{\circ}$. 
The differences at all depth targets were in all cases much less than the formal uncertainties, and the small differences within formal uncertainties were neither systematically positive nor negative across all latitudes, including the Equator and 60$^{\circ}$ N latitude shown in Figure~\ref{fig:Fig12}. 

%\subsection{Time-Distance Helioseismology}
%\label{sec_TD}

\begin{figure}[!htbp]
    \centering
    \includegraphics[width=0.95\textwidth]{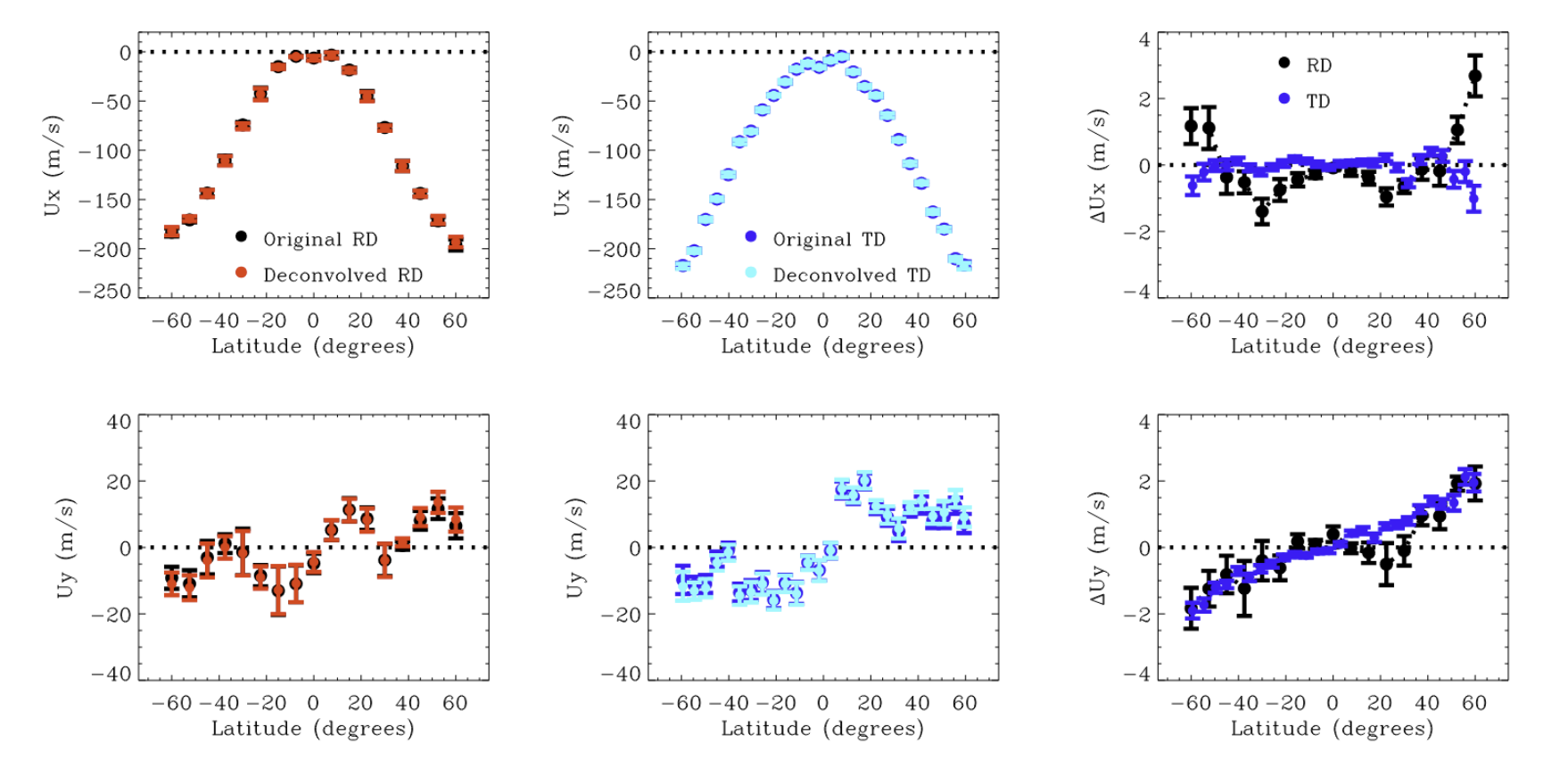}
    \caption{{\it Upper:} 
    Differential rotation inferred at $1-3$ Mm depth. The left and middle panels show results from original and deconvolved HMI Dopplergrams using ring-diagram (RD) and time-distance (TD) analyses, respectively. The right panel displays the differences between original and deconvolved results for both methods.
    {\it Lower:} Same as the upper row but for the meridional-flow profiles (before the center-to-limb effect is removed) obtained at the same depth. The results are from an average over a five day period and with spatial resolution of 15$^{\circ}$  per pixel for ring-diagram analysis and 1.2$^{\circ}$ per pixel for time-distance analysis.}
    \label{fig:Fig13}
\end{figure}

 Figure~\ref{fig:Fig13} shows results from both time-distance and ring-diagram analysis methods as a function of latitude along the central meridian at a given depth.
For the near-surface rotation, the two sets of measurements show very similar results, but it can also be found that the rotation measured from the deconvolved Doppler data is slightly faster than that measured from the original Doppler data by an average of approximately 0.5\,m\,s$^{-1}$ (see Figure~\ref{fig:Fig13}).  This difference does not have a clear trend and may be negligible in the differential rotation measurements that typically have a magnitude of about 200\,m\,s$^{-1}$. 
However, the difference in the meridional-flow profiles is much more systematic and non-negligible.   Figure~\ref{fig:Fig13} show that the deconvolution of Doppler velocities introduces a systematic latitude-dependent difference relative to measurements using the original Doppler velocities. 
This difference has an order of magnitude of 2\,m\,s$^{-1}$, and looks similar to the systematic helioseismic center-to-limb effect discussed by various authors \citep{Zhao:2012apjl, Baldner:2012ApJL, Chen:2018ApJ}.  
Note that this systematic difference changes with increasing depth, and becomes negligible in the deeper interior.
Note that both of the meridional-flow profiles do not have their respective center-to-limb effects removed, because the purpose of this comparison is to show how the deconvolution impacts measurements.  As a reference to the readers, our data experiments show that the two meridional-flow profiles still differ similarly after their respective center-to-limb effects are removed.

\begin{figure}[h]
    \centering
    \includegraphics[width=0.60\textwidth]{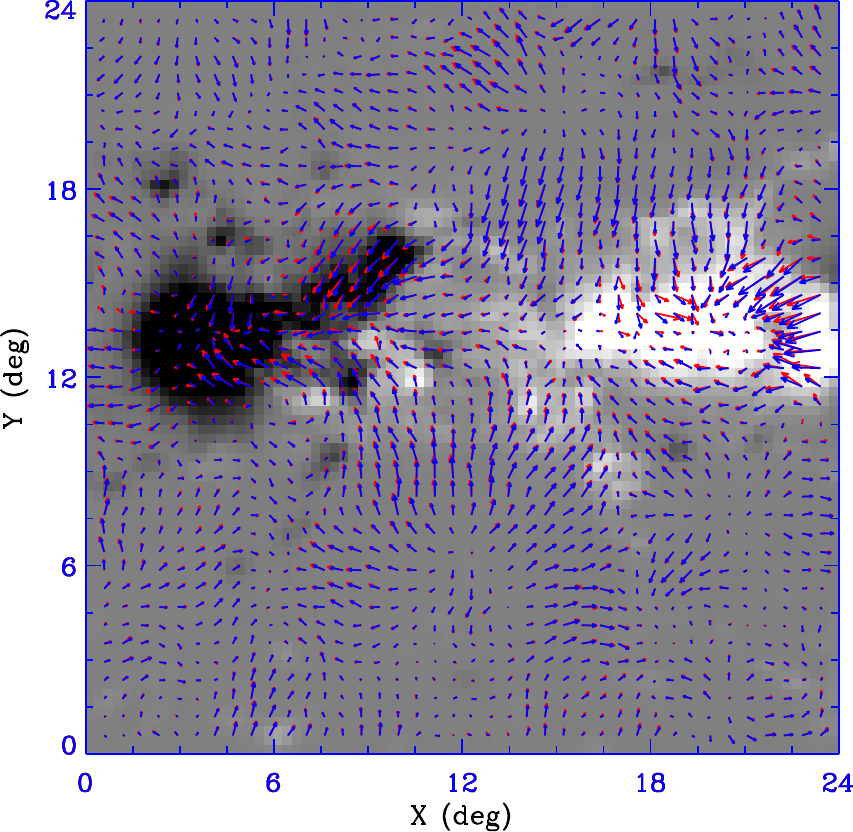}
    \caption{Comparison of the near-surface flow fields obtained at the depth of $0-1$ Mm around an active region from an eight-hour time period with a spatial resolution of 0.12$^{\circ}$. Red arrows are flows inferred from the original Dopplergrams and the blue arrows are from the deconvolved Dopplergrams. The longest arrow in the plot represents a speed of 450\,m\,s$^{-1}$. }
    \label{fig:Fig14}
\end{figure}

Figure~\ref{fig:Fig14} shows a comparison of the subsurface flow fields
obtained by time-distance analysis and 
inferred from the original and deconvolved set of Dopplergrams, which include a sunspot in the field of view.
As can be seen, the supergranular flow patterns from both sets of the data are very similar to each other, with similar flow magnitudes and flow directions. 
However, significant differences are visible in the sunspot region, with slightly different flow directions.
The magnitude of the differences are still considered small, likely comparable to the measurement uncertainties in the active regions, which are around 50\,m\,s$^{-1}$ at this depth \citep{Zhao:2012SP}. 

%\section{Data Availability}

%Beginning in 2018, one set of observables, both 45-second and 720-second data, were provided every day at a T\_REC of 19:00 UT. Other data are available from past requests, notably NOAA 11158, 13179, 13664 and 13668. See the coverage charts at:
%\begin{itemize}
%\item http://jsoc.stanford.edu/data/dcon45.html for the 45-second data coverage,
%  \item http://jsoc.stanford.edu/data/dcon720s.html for the 720-second data coverage, and 
%  \item http://jsoc.stanford.edu/data/cont\_dcon.html for the true continuum.   
%\end{itemize}

%Additional stray-light corrected data can be requested by individual users and will be processed and provided by the HMI team as resources allow.

\section{Conclusions}\label{ref:sec5}
HMI data are used extensively by solar researchers the world over. The science goals that this data upgrade enable include the following.
\begin{itemize}
\item The scattered light correction enhances visual clarity, resulting in more striking and detailed imagery, which can be useful for scientific purposes as well as for presentations to the public. One example of the enhanced visual clarity can be seen in the video of flux emergence of AR 13179 found \href{https://vimeo.com/1096037326}{here}. 

\item The correction improves quality of solar irradiance reconstruction by providing full disk, daily data with higher photometric and magnetic contrast accuracy, i.e. improved modeling results produced by codes such as SATIRE-S \citep{yeo:2014}. \citet{criscuoli:2017} used HMI data corrected by scattered light for such purposes. After identifing faculae and plage based on proximity to active regions in this upgraded data, they found a marked increase in network brightness with higher center-to-limb variations than faculae, consistent with findings from high resolution data. They show that using proximity to active regions, in addition to field strength, for faculae and network identification and characterizing their photometric contrast with the new characteristics decreases the estimated TSI for cycle 24 by 11\% .  
\item The correction increases accuracy of co-alignment between HMI and higher-spatial resolution instruments such as IRIS \citep{depontieu:2014}, DKIST \citep{rimmele:2020}, IBIS \citep{cavallini:2006}, Swedish Solar Tower CRISP \citep{scharmer:2003}, Hinode SOT SP \citep{tsuneta:2008}, Solar Orbiter PHI \citep{solanki:2020}, etc., that use HMI for the larger field of view context. HMI data can inform researchers about magnetic connectivity and dynamics of nearby regions. Positioning high-resolution observations into a larger field-of-view map from HMI assists in interpretation of dynamics. \citet{houston:2018} used scattered light corrected $I_c$ and $B$ data for an inter-instrumental comparison for 2016 July 14 data which resulted in a exceptional view of magnetoacoustic shock phenomena and the impact these shocks have on the surrounding magnetically dominated plasma.

\item It decreases tracking errors by providing larger intensity contrasts of features. For example, the determination of sunspot rotation rates using algorithms similar to \citet{brown:2021} have reduced errors using stray-light-corrected data, thus providing greater confidence in results. 

\item  Improvements in field strength measurements are found for pores, plage and other  magnetic features since the original HMI field strengths were contaminated by stray light. The corrected data have values in closer agreement with the Hinode SP values, see Figure 13 in \citet{beck:2025}, but it is argued that a scaling curve could bring the HMI field strength values into better agreement with Hinode SP \citep{tsuneta:2008} data as the magnetic filling fraction is still not fit in the stray-light-corrected data. 

\item The scattered-light correction does significantly alter the direct Doppler velocities measured, partially correcting for the convective blueshift, and potentially improving analysis of direct surface Doppler measurements, see Figures~\ref{fig:Fig7} and  \ref{fig:Fig8}.

\item 
In general, the scattered light correction introduces negligible effects on local helioseismology analysis for determining long-term rotation speeds, averaged in time over five days and spatially over many heliographic degrees, see Figures~\ref{fig:Fig12} and \ref{fig:Fig13}. Also, a systematic shift resembling the helioseismic center-to-limb effect is found in the near-surface meridional-flow measurements on the order of 2 m s$^{-1}$, see Figure~\ref{fig:Fig13}, lower panels. 

\item For local helioseismology analysis on shorter timescales and with higher spatial resolution, the sub-surface flow field differences are found to be negligible in quiet-Sun regions but show significant differences in and near sunspot regions, see Figure~\ref{fig:Fig14}.

\end{itemize}

Currently, the HMI team is producing data corrected for scattered light once a day at 19:00 UT, but other time periods have been processed as requested by users and are available; the links to the coverage charts can be found in the Appendix. The HMI team encourages requests for the data at other times. Please email a team member listed under \textit{contacts} at the JSOC web page \href{http://jsoc.stanford.edu}{http://jsoc.stanford.edu}. If a full recalibration of the HMI data were to occur in the future, for example, to correct for the 24-hour variation, then we recommend that the scattered light correction is applied to all data at that time. 

\acknowledgments
This work was supported by NASA contracts HiDEE 80NSSC18K0380 and NAS5-02139 (HMI) to Stanford University. 

\clearpage
\bibliographystyle{apj}
\bibliography{SL-Norton}
\appendix
%\section{Data Availability}
\setcounter{table}{0}
\renewcommand{\thetable}{A\arabic{table}}
The HMI data described in this paper are heterogeneous and somewhat complex; for convenience we describe the relevant characteristics in this appendix, as well as how these data can be found and accessed. All HMI data are captured and processed at the Joint Science Operations Center (JSOC) located at Stanford University. Data products are organized into series, identified by the instrument and then the data product name, e.g. hmi.B\_720s. HMI data can be queried and accessed through the JSOC website\footnote{\href{http://jsoc.stanford.edu}{http://jsoc.stanford.edu}}. 

At the lowest level, actual images captured by the HMI instrument are often called filtergrams --- single exposures at a given wavelength bandpass and polarization state. The combination of wavelength tuning and polarization selection is encoded in the parameter Filtergram ID, or FID. Raw HMI images are termed level 0 and are not used in this work. Images that have been corrected for known CCD systematics are termed level 1, and have the series name hmi.lev1. The data in hmi.lev1 are generally not accessible outside the JSOC except upon request, but special filtergrams including daily calibrations are available in the series hmi.lev1\_cal.

The lowest level of science data products are termed level 1.5, and include full disk images at 45 second cadence and 720 second cadence. Products include photospheric velocity along the line of sight ($V$), continuum images $I_c$, magnetic field along the line of sight ($M$), magnetic vector field (720s only; $B$), line width ($Lw$), and line depth ($Ld$). Stokes $I, Q, U, V$ are also produced at 720s cadence as an intermediate product. Higher level science products include the Space  Weather Active Region Patch products (SHARPs) which are cutouts around active regions, and helioseismic subsurface flow maps. A detailed description of the procedures involved in HMI data processing can be found in \citet{couvidat:2016}. 

Data products are PSF-deconvolved at one of two stages: either the level 1 filtergrams, or the Stokes images. Products using the former data have `\_dcon' appended to the series name; products based on the deconvolved Stokes images have `\_dconS' appended to the series name. The names of relevant data product series names, and their PSF-deconvolved equivalents, are shown in Table~\ref{tab:A1}.

\begin{table*}[hb]
\caption{Names of Original \& Upgraded Data Products}
\centering
\begin{tabular}{ll}
\hline  
Original & PSF Corrected \\
\hline
\hline
\textbf{45-second cadence data} &    \\
\hline
hmi.Ic\_45s & hmi.Ic\_45s\_dcon \\
hmi.M\_45s & hmi.M\_45s\_dcon \\
hmi.V\_45s & hmi.V\_45s\_dcon \\
hmi.Ld\_45s & hmi.Ld\_45s\_dcon \\
hmi.Lw\_45s & hmi.Lw\_45s\_dcon \\
\hline 
\textbf{720-second cadence data} &    \\
\hline
hmi.Ic\_720s & hmi.Ic\_720s\_dconS \\
hmi.M\_720s & hmi.M\_720s\_dconS \\
hmi.V\_720s & hmi.V\_720s\_dconS \\
hmi.Lw\_720s & hmi.Lw\_720s\_dconS \\
hmi.Ld\_720s & hmi.Ld\_720s\_dconS \\
hmi.B\_720s  & hmi.B\_720s\_dconS (multiple segments) \\
hmi.sharp\_720s & hmi.sharp\_720s\_dconS (multiple segments)\\
hmi.sharp\_cea\_720s & hmi.sharp\_cea\_720s\_dconS (multiple segments)\\
\hline
\textbf{True continuum}   &(tuned 0.345~{\AA}~from line center)    \\
\hline
hmi.lev1~[FID=10001] & hmi.cont\_dcon  \\
\hline
\end{tabular}
\label{tab:A1}
\end{table*}

During the limb-fitting procedure, the image center is recorded as the CRPIX1 and CRPIX2 keywords. RSUN\_OBS is a known quantity determined from an ephemeris value that represents the radius of the Sun for the spacecraft position. RSUN\_LF is the fit value returned from the limb fit program. The image pixel scale in the X and Y direction are recorded as CDELT1 and CDELT2 and are determined by dividing RSUN\_OBS by RSUN\_LF.  If RSUN\_LF is slightly smaller (larger) between the dcon and dconS data, then the pixel scale increases (decreases) by a small amount. The value of RSUN\_LF is not included in Dopplergram or magnetic field products, but changes in pixel scale are apparent in the CRPIX1 and CRPIX2 keywords A full description of the algorithms used to calculate the HMI values can be found in the Calculation and Measurement Algorithm Document (CMAD) for HMI, which will be placed in the Stanford Digital Repository \href{https://purl.stanford.edu/gz366yd5962}{https://purl.stanford.edu/gz366yd5962}.

Regarding data availability, beginning in 2018, one set of observables, both 45-second and 720-second data, were provided every day at a T\_REC of 19:00 UT. Other data are available from past requests, notably NOAA 11158, 13179, 13664 and 13668. 

See the \textbf{coverage charts} at:
\begin{itemize}
\item 45-second data: \href{http://jsoc.stanford.edu/data/dcon45.html}{http://jsoc.stanford.edu/data/dcon45.html},
  \item 720-second data: \href{http://jsoc.stanford.edu/data/dcon720s.html}{http://jsoc.stanford.edu/data/dcon720s.html}, and 
  \item True continuum: \href{http://jsoc.stanford.edu/data/cont_dcon.html}{http://jsoc.stanford.edu/data/cont\_dcon.html}.
\end{itemize}

Additional stray-light corrected data can be requested by individual users and will be processed and provided by the HMI team as resources allow.
\end{document}